\documentstyle[epsfig,aps]{revtex}
\begin{document}
\draft

\title{Trapped condensates of atoms with dipole interactions}
\author{S. Yi and L. You}
\address{School of Physics, Georgia Institute of Technology,
Atlanta, GA 30332-0430}
\date{\today}
\maketitle

\begin{abstract}
We discuss in detail properties of trapped atomic
condensates with anisotropic dipole interactions.
A practical procedure for constructing anisotropic low energy
pseudo potentials is proposed and justified by the agreement
with results of numerical multi-channel calculations.
The time dependent variational method is adapted to
reveal several interesting features observed in numerical
solutions of condensate wave function.
Collective low energy shape oscillations and their
stability inside electric fields are investigated.
Our results shed new light into macroscopic coherence
properties of interacting quantum degenerate atomic gases.
\end{abstract}



\section{introduction}
The recent success in atomic Bose-Einstein condensation (BEC)
\cite{bec,mit,rice} has stimulated great research activities into
trapped quantum gases \cite{Edwards}. To a remarkable
degree, a single condensate wave function of the macroscopically
occupied ground state, described by the nonlinear Schr\"odinger
equation (NLSE) \cite{nlse}, captures all essential features of its
coherence properties \cite{Stringari}. In fact, one of
the key diagnosis features for BEC, the reversed aspect ratio
of a free expanding condensate, is described purely by the
condensate wave function \cite{dum,kagan1}.
In the standard treatment for the condensate wave function
of interacting atoms, realistic inter-atomic potential
$V(\vec R)$ is often not directly used. Instead,
a contact pseudo potential, $u_0\delta(\vec R)$, obtained under the
so-called shape independent approximation (SIA) \cite{yang} is used.
Such an idealization results in tremendous simplification,
yet to date, SIA has worked remarkably well as verified
by both theoretical calculations and experimental
observations \cite{Edwards,glauber,esry}.

Currently available degenerate quantum gases are cold and dilute,
the interaction is therefore dominated by s-wave collisions,
described by a single atomic parameter: $a_{\rm sc}$, the s-wave
{\it scattering length}, if the inter-atomic potential is
isotropic and short ranged (decaying fast than $-1/R^3$
asymptotically). The complete scattering amplitude is then
isotropic and energy-independent, given by $f(\vec k,\vec
k')=-4\pi a_{\rm sc}$ for collisions of incoming momentum $\vec
k$ state scattering into $\vec k'$.

One of the attractive features of atomic degenerate
gases lies at effective means for control of
the atom-atom interaction \cite{gora,verhaar,eite}.
Indeed, very recently several groups have successfully
implemented {\it Feschbach resonance} \cite{frec,fret},
thus enabling a control knob on $a_{\rm sc}$ through the
changing of an external magnetic field.
Other physical mechanisms also exist for modifying
atom-atom interactions, e.g. the {\it shape resonance}
as proposed in \cite{mm}.
In an external electric field,
inter atomic potential is modified by the addition of
an anisotropic (induced) dipole interaction.

Although anisotropically interacting fermi system has been an
important area of study, e.g. liquid $^3$He \cite{leggett} and
d-wave paired high $T_c$ superconductors \cite{davis}. Its
bosonic counterpart has not been studied in great detail. In
particular, we are not aware of any systematic approach for
constructing an anisotropic pseudo potential \cite{yang}.

For bosonic systems, another related topic is
the condensate stability.
Under the SIA, the scattering length takes a
positive or negative value,
corresponding to repulsive or attractive interactions.
When $a_{\rm sc}<0$ occurs, self-interaction leads to
a collapse of BEC in dimensions higher than 1 \cite{kb},
thus the resulting condensate
is limited by a critical number of particles \cite{bradley}.
Anisotropic dipole interactions, on the other hand,
are more complicated as both attractive and repulsive
interactions arise along different directions. We note
that several recent investigations have studied
efforts of non-local interactions on condensate stability
\cite{ml,parola}.

In this paper, we study properties of trapped BEC of
atoms with dipole interactions \cite{yi},
arising from either external electric field (induced)
or permanent magnetic moments \cite{goral}.
We propose a practical method for constructing
anisotropic pseudo potentials that can also
be extended for investigation of
polar molecular BEC \cite{doyle,stwalley}.
This paper is organized as following. We first briefly review
the usual pseudo-potential approximation under the SIA.
In Sec. II we describe and justify in detail a
procedure for constructing effective low energy
pseudo-potentials of anisotropic interactions.
In Sec. III we provide our numerical procedure for
solving the NLSE with anisotropic
dipole interactions. Particular emphasis is put on the
careful treatment of the singular origin of
dipole interactions. We also present and discuss
results from selected numerical calculations.
To explain the stability region
as well as the interesting aspect ratios observed from
our numerical calculations, we perform in Sec. IV
an analytic time dependent variational calculation.
We compare the results obtained with direct
numerical solutions of NLSE.
Finally we conclude.

\section{formulation}
For $N$ trapped spinless bosonic atoms in a potential
$V_t(\vec r)$, the second quantized Hamiltonian is given by
\begin{eqnarray}
{\cal H}&=& \int\! d\vec r\,\hat\Psi^{\dag}(\vec r)
\left[-\frac{\hbar^2}{2M}\nabla^2+V_{t}(\vec r)
-\mu\right]\hat\Psi(\vec r) \nonumber\\
&+&\frac{1}{2}\int\!
d\vec r\int\! d\vec r' \hat\Psi^{\dag}(\vec r)\hat\Psi^{\dag}(\vec
r') V(\vec r-\vec r')\hat\Psi(\vec r')\hat\Psi(\vec r), \label{h}
\end {eqnarray}
where $\hat\Psi(\vec r)$ and $\hat\Psi^{\dag}(\vec r)$ are atomic
(bosonic) annihilation and creation fields. The chemical potential
$\mu$ guarantees the atomic number $\hat N= \int d\vec
r\,\hat\Psi^{\dag}(\vec r)\hat\Psi(\vec r)$ conservation.

The bare potential $V(\vec R)$ in (\ref{h}) needs to be
renormalized for a meaningful perturbation calculation \cite{yang}.
The usual treatment is based an effective
interaction obtained by a resummation of certain classes of
interaction diagrams \cite{Galits,Beliaev,Brueckner}.
Physically the SIA can be viewed as a valid low energy
and low density renormalization scheme, one simply
replaces the bare potential $V(\vec R)$
by a pseudo potential $u_0\delta(\vec R)$ whose first
order Born scattering amplitude reproduces the complete scattering
amplitude ($-a_{\rm sc}$). This gives $u_0=4\pi\hbar^2 a_{\rm
sc}/M$.

When an electric field is introduced along the positive z axis, an
additional term
\begin{eqnarray}
V_{E}(\vec R)&&=u_2{Y_{20}(\hat R)\over R^3}, \label{ve}
\end{eqnarray}
appears in the atom-atom interaction \cite{mm}, where
$u_2=-4\sqrt{(\pi/5)}\,\alpha(0)\alpha^*(0){\cal E}^2$.
$\alpha(0)$ is the atomic polarizability, and ${\cal E}$
denotes the electric field strength. As was shown before
\cite{mm}, this modification results in a completely
new form for the low-energy scattering amplitude
\begin{eqnarray}
f(\vec k,\vec k')\Big|_{k=k'\to 0}=4\pi\sum_{lm,l'm'}
t_{lm}^{l'm'}({\cal E})Y_{lm}^*({\hat k})Y_{l'm'}({\hat k'}),
\label{cf}
\end{eqnarray}
with $t_{lm}^{l'm'}({\cal E})$ the reduced T-matrix elements. They
are all energy independent and act as generalized scattering
lengths. The anisotropic $V_{E}$ causes the dependence on both
incident and scattered directions: $\hat k$ and $\hat k'=\hat R$.

We therefore propose a general (energy-independent)
anisotropic pseudo potential constructed
according to
\begin{eqnarray}
V_{\rm eff}(\vec R)=u_0\delta(\vec R) +\sum_{l_1>0, m_1}
\gamma_{l_1m_1} {Y_{l_1m_1}(\hat R)\over R^3}, \label{veff}
\end{eqnarray}
whose first Born amplitude is
\begin{eqnarray}
f_{\rm Born}(\vec k,\vec k') =-(4\pi)^2 a_{\rm sc}Y_{00}^*(\hat
k)Y_{00}(\hat k') -{M\over 4\pi\hbar^2}\sum_{l_1 m_1}
\gamma_{l_1m_1} (4\pi)^2 \sum_{lm} \sum_{l'm'}{\cal
T}_{lm}^{l'm'}(l_1,m_1) Y_{lm}^*(\hat k)Y_{l'm'}(\hat k'),
\label{bf}
\end{eqnarray}
with $ {\cal T}_{lm}^{l'm'}(l_1,m_1)=(i)^{l+l'}{\cal R}_{l}^{l'}
I_{lm}^{l'm'}(l_1,m_1).$ Both
\begin{eqnarray}
I_{lm}^{l'm'}(l_1m_1)&&=\langle
Y_{l'm'}|Y_{l_1m_1}|Y_{lm}\rangle\nonumber\\
&&=(-1)^m\sqrt{(2l+1)(2l'+1)(2l_1+1)\over 4\pi}\left(
\begin{array}{ccc} l & l' & l_1\\ -m & m' & m_1\end{array} \right)
\left( \begin{array}{ccc} l & l' & l_1\\ 0 & 0 & 0\end{array}
\right),
\end{eqnarray}
and
\begin{eqnarray}
{\cal R}_{l}^{l'} &&=\int_0^{\infty} d R\,{1\over
R}j_l(kR)j_{l'}(k'R)\nonumber\\ &&={\pi\over 8}\eta^l
{\Gamma({l+l'\over 2})\over \Gamma({3+l'-l\over
2})\Gamma(l+{3\over 2})} \ _2F_1\left ({-1-l'+l\over 2},{l+l'\over
2},l+{3\over 2}; \eta^2\right),
\end{eqnarray}
can be computed analytically. The $1/R^3$ form in Eq. (\ref{veff})
assures all ${\cal R}_{l}^{l'}$ to be $k=k'$ independent
(easily seen by a change of variable to $x=kR$ in the integral).
Putting
\begin{eqnarray}
f_{\rm Born}(\vec k,\vec k')=f(\vec k,\vec k'),
\end{eqnarray}
i.e. requiring the Bohn amplitude from the pseudo potential
Eq. (\ref{veff}) to be the same as the numerically computed
value $f(\vec k,\vec k')$,
one can solve for all $\gamma_{l_1m_1}({\cal E})$ from
known $t_{lm}^{l'm'}({\cal E})$ \cite{mm,mm2}.
This reduces to a set of (under determined) linear equations
\begin{eqnarray}
-{M\over 4\pi \hbar^2}\sum_{l_1 m_1} \gamma_{l_1m_1} (4\pi) {\cal
T}_{lm}^{l'm'}(l_1,m_1)\equiv t_{lm}^{l'm'},
\label{eall}
\end{eqnarray}
for all ($lm$) and ($l'm'$) with $l,l'\ne 0$, and separately
$a_{\rm sc}({\cal E})=-t_{00}^{00}({\cal E})$.

\begin{figure}[h]
\centerline{\epsfig{file=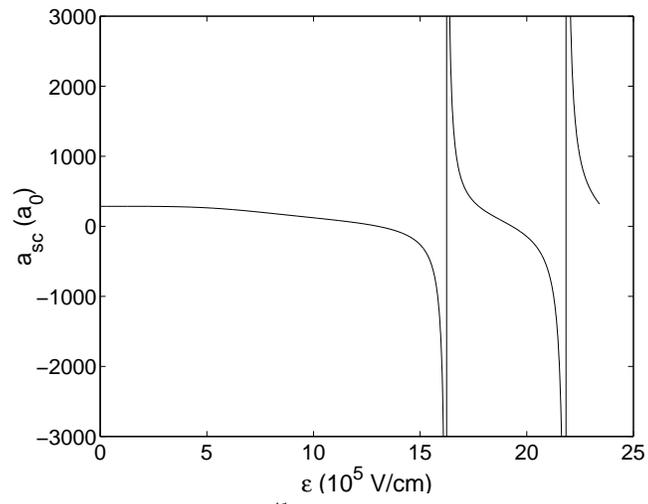,width=8.5cm}} \caption{$a_{\rm
sc}$ v.s. ${\cal E}$ for $^{41}$K in electron spin triplet
state.} \label{fasc}
\end{figure}

Considerable simplification arises further for bosons (fermions)
as only even (odd) $(l,l')$ terms are needed to match in
(\ref{eall}). Figure \ref{fasc} displays the result of field
dependent $a_{\rm sc}({\cal E})$ $^{41}$K atoms in the triplet
electron spin state. Note the spikes of shape resonances. The
Born amplitude for the dipole term $V_E$ is
\begin{eqnarray}
f_{\rm Born}(\vec k,\vec k') &&=u_2{M\over 4\pi \hbar^2}
(4\pi)^2{\cal T}_{00}^{20} \sum_{lm,l'm'}\overline{\cal
T}_{lm}^{l'm'} Y_{lm}^*(\hat k)Y_{l'm'}(\hat k'),
\end{eqnarray}
with ${\cal T}_{00}^{20}=-0.023508$. $\overline {\cal
T}_{lm}^{l'm'}={\cal T}_{lm}^{l'm'}(2,0)/{\cal T}_{00}^{20}$ are
independent of electric field ${\cal E}$ within the perturbative
Born approximation as tabulated below.

\begin{table}
\caption{$\bar{\cal T}_{lm}^{l'm'}= t_{lm}^{l'm'}({\cal
E})/t_{00}^{20}({\cal E})$ for small $(l,l')$.}
\begin{tabular}{c|ccccc}
$(lm),(l'm')$&(00)&(20)&(40)&(60)&(80)\\ \hline\\[-9pt] (00)&0&1&0&0&0\\
(20)&1&-0.63889&0.14287&0&0\\(40)&0&0.14287&-0.17420&0.05637&0\\
(60)&0&0&0.05637&-0.08131&0.03008\\(80)&0&0&0&0.03008&-0.04707
\end{tabular}
\label{table1}
\end{table}

We find that away from regions of {\it shape resonances} to be
discussed elsewhere \cite{mm2}, all numerically
computed $t_{lm}^{l'm'}({\cal E})$ values, large
enough to justify their inclusions, are actually all
proportional to ${\cal E}^2$.
Thus, we could rewrite $f(\vec k,\vec k')$ as
\begin{eqnarray}
f(\vec k,\vec k')=(4\pi)t_{00}^{20}({\cal E})\sum_{lm,l'm'}
\overline {t}_{lm}^{l'm'}Y_{lm}^*({\hat k})Y_{l'm'}({\hat k'}),
\end{eqnarray}
with scaled quantities
$\overline {t}_{lm}^{l'm'}= t_{lm}^{l'm'}({\cal
E})/t_{00}^{20}({\cal E})$ now all being constants.
We have since computed (numerically) for several
alkali metal isotopes, our results for $\overline
{t}_{lm}^{l'm'}$ are tabulated below.
\begin{table}
\caption{$\overline {t}_{lm}^{l'm'}$ for $^7$Li.}
\begin{tabular}{c|ccccc}
$(lm),(l'm')$&(00)&(20)&(40)&(60)&(80)\\ \hline\\[-9pt]
(00)&0&1.0&0.0&0.0&0.0\\ (20)&1.0&-0.63&0.14&0.0
&0.0\\(40)&0.0&0.14&-0.17&0.057&0.0\\
(60)&0.0&0.0
&0.057&-0.080&0.031\\(80)&0.0&0.0&0.0&0.031&-0.044
\end{tabular}
\label{tabli7}
\end{table}

\begin{table}
\caption{$\overline {t}_{lm}^{l'm'}$ for $^{39}$K.}
\begin{tabular}{c|ccccc}
$(lm),(l'm')$&(00)&(20)&(40)&(60)&(80)\\ \hline\\[-9pt]
(00)&0&1.0&0.0&0.0&0.0\\ (20)&1.0&-0.64&0.14&0.0
&0.0\\(40)&0.0&0.15&-0.17&0.085&0.0\\
(60)&0.0&0.0
&0.085&-0.127&0.047\\(80)&0.0&0.0&0.0&0.047&-0.074
\end{tabular}
\label{tabk39}
\end{table}

\begin{table}
\caption{$\overline {t}_{lm}^{l'm'}$ for $^{41}$K.}
\begin{tabular}{c|ccccc}
$(lm),(l'm')$&(00)&(20)&(40)&(60)&(80)\\ \hline\\[-9pt]
(00)&0&1.0&0.0&0.0&0.0\\ (20)&1.0&-0.64&0.14&0.0
&0.0\\(40)&0.0&0.14&-0.17&0.057&0.0\\
(60)&0.0&0.0
&0.057&-0.081&0.030\\(80)&0.0&0.0&0.0&0.030&-0.047
\end{tabular}
\label{tabk41}
\end{table}

\begin{table}
\caption{$\overline {t}_{lm}^{l'm'}$ for $^{85}$Rb.}
\begin{tabular}{c|ccccc}
$(lm),(l'm')$&(00)&(20)&(40)&(60)&(80)\\ \hline\\[-9pt]
(00)&0&1.0&0.0&0.0&0.0\\ (20)&1.0&-0.64&0.14&0.0
&0.0\\(40)&0.0&0.14&-0.17&0.056&0.0\\
(60)&0.0&0.0
&0.056&-0.081&0.030\\(80)&0.0&0.0&0.0&0.030&-0.047
\end{tabular}
\label{tabrb85}
\end{table}

\begin{table}
\caption{$\overline {t}_{lm}^{l'm'}$ for $^{87}$Rb.}
\begin{tabular}{c|ccccc}
$(lm),(l'm')$&(00)&(20)&(40)&(60)&(80)\\ \hline\\[-9pt]
(00)&0&1.0&0.0&0.0&0.0\\ (20)&1.0&-0.64&0.15&0.0
&0.0\\(40)&0.0&0.15&-0.17&0.056&0.0\\
(60)&0.0&0.0
&0.056&-0.079&0.032\\(80)&0.0&0.0&0.0&0.032&-0.045
\end{tabular}
\label{tabrb87}
\end{table}

The agreement between the first order Born approximation and the
multi-channel scattering calculations is remarkable. We estimate
the numerical scattering results to be accurate to a few percent
(except for $^{39}$K), independent of atoms being bosons (even
$l,l'$) or fermions (odd $l,l'$). Only bosonic results are being
considered in this paper. This is displayed by noticing the
agreement between Table \ref{table1} ($\sim$ a few per cent) with
Table \ref{tabli7}-\ref{tabrb87}. This interesting observation
applies for all bosonic alkali triplet states: $^7$Li,
$^{39,41}$K, and $^{85,87}$Rb, for up to a field strength of
$3\times 10^6$ (V/cm) \cite{mm,mm2} computed by us. Physically,
this implies the effect of $V_E$ is perturbative when ${\cal E}$
remains small (in a.u.). For the convenience of further
discussions, we tabulate polarizabilities of selected atoms in
Table \ref{tabpolar}.
\begin{table}
\caption{Atomic polarizabilities in (a.u.).}
\begin{tabular}{ccccccc}
    &H & Li & Na & K & Rb \\
\tableline\\[-9pt]
 $\alpha(0)$ & 4.5 & 159.2 & 162 & 292.8& 319.2
\end{tabular}
\label{tabpolar}
\end{table}
What is remarkable is the fact that ${\cal T}_{00}^{20}({\cal E})$ and
$t_{00}^{20}({\cal E})$ also agree in absolute values \cite{mm}
except for a slight difference (1-6\%). They are calculate
below and tabulated in \ref{tabli}.

\begin{eqnarray}
u_2{M\over 4\pi\hbar^2} (4\pi)^2{\cal T}_{00}^{20}
&&=-16\pi\sqrt{\pi\over 5}{\cal T}_{00}^{20}\alpha^2{\cal
E}^2{M\over \hbar^2}\nonumber\\&&=1718\times\overline{M}
\overline{\alpha}^2\overline{\cal E}^2a_0,
\end{eqnarray}
where all overlined quantities are in atomic units (a.u.).

\begin{table}
\caption{Comparison of numerical values of $u_2{M\over
4\pi\hbar^2} (4\pi)^2{\cal T}_{00}^{20}$ with
$-(4\pi)t_{00}^{20}({\cal E})$, the cause of the slight
difference is unclear but within numerical errors.}
\begin{tabular}{c|l|l}
 \    &Born  & Multichannel \\[3pt]
atom&$u_2{M\over 4\pi\hbar^2} (4\pi)^2{\cal
T}_{00}^{20}(a_0)$\ \ &$-(4\pi)t_{00}^{20}({\cal E})(a_0)$\\
\hline\\[-9pt]
$^7$Li&$3.040\times 10^8 {\overline {\cal E}}^2$&$3.238\times
10^8{\overline {\cal E}}^2$
\\$^{39}$K&$5.713\times 10^9 {\overline {\cal
E}}^2$&$5.699\times 10^9{\overline {\cal E}}^2$
\\$^{41}$K&$6.006\times 10^9 {\overline {\cal
E}}^2$&$5.99\times 10^9{\overline {\cal E}}^2$
\\$^{85}$Rb&$1.486\times 10^{10} {\overline {\cal
E}}^2$&$1.474\times 10^{10}{\overline {\cal E}}^2$
\\$^{87}$Rb&$1.495\times 10^{10} {\overline {\cal
E}}^2$&$1.512\times 10^{10}{\overline {\cal E}}^2$
\end{tabular}
\label{tabli}
\end{table}

An important parameter in our discussion is the ratio between
$u_2$ and $u_0$. Since the results from first order Born
approximation and the multichannel calculations are about
the same, one can write this ratio as
\begin{eqnarray}
c({\cal E})=-{u_2\over u_0}=\gamma{\cal E}^2,
\end{eqnarray}
where $\gamma\simeq 1.748\times 10^{-17}\overline{\alpha}^2\overline
M/\overline a_{\rm sc}$ and ${\cal E}$ is in unit of V/cm. The
$\gamma$ values for selected atoma are tabulated
in Table \ref{tabpolar1}.
\begin{table}
\caption{Values of parameter $\gamma$ for selected atoms
with their field free (${\cal E}=0$) scattering lengths.}
\begin{tabular}{ccccccc}
    & H & $^7$Li & $^{23}$Na & $^{41}$K & $^{87}$Rb \\
\tableline\\[-9pt] $\gamma\,(\times 10^{-13})$ & $2.9\times
10^{-3}$ & $-1.1$ & $2.0$&$2.2$& $15.0$
\end{tabular}
\label{tabpolar1}
\end{table}

Most atoms possess permanent magnetic dipoles.
With alkali metals, the magnetic dipole mainly
originates from valance electron spin, typically
measured in units of Bohr magneton. It is therefore
interesting to compare electric dipole interactions with
magnetic dipole interactions. The dipole interaction strength
between atoms of a permanent magnetic dipole $\mu$ is
\begin{eqnarray}
\mu^2&&={\overline\mu}^2\mu_B^2\nonumber\\&&=1.331\times
10^{-5}{\overline\mu}^2e^2a_0^2,
\end{eqnarray}
with $\mu_B$ the unit of Bohr magneton. For induced
electric dipoles, the interaction strength is
\begin{eqnarray}
\alpha^2{\cal E}^2=
{\overline\alpha}^2{\overline{\cal E}}^2e^2a_0^2.
\end{eqnarray}
A typical heavy alkali atom has ${\overline \alpha}\sim 200$,
thus for which  a 1 ($\mu_B$) magnetic moment corresponds to an
effective electric field of $3.3\times 10^{-5}$ (a.u.), or
$1.7\times 10^5$ (V/cm). Atoms with large magnetic
moments effectively simulates induced dipole interactions
at a high value of equivalent electric field \cite{goral}.

As can be concluded from comparing listed data in
all tables, we can approximate Eq. (\ref{veff}) by
keeping only the $l_1=2$, $m_1=0$ term to achieve
a satisfactory level of accuracy. Thus away from
{\it shape resonances} we take
\begin{eqnarray}
V_{\rm eff}(\vec R)=u_0\delta(\vec R)+u_2 Y_{20}(\hat R)/R^3,
\end{eqnarray}
where $u_0={4\pi\hbar^2 \over M} a_{\rm sc}({\cal E})$
and $u_2=-c({\cal E})u_0$.
The Hamiltonian (\ref{h}) then becomes
\begin{eqnarray}
{\cal H}&=& \int\! d\vec r\,\hat\Psi^{\dag}(\vec r)
\left[-\frac{\hbar^2}{2M}\nabla^2+V_{t}(\vec r)
-\mu\right]\hat\Psi(\vec r) \nonumber\\
&+&\frac{u_0}{2}\int\!
d\vec r\, \hat\Psi^{\dag}(\vec r)\hat\Psi^{\dag}(\vec r)
\hat\Psi(\vec r)\hat\Psi(\vec r)\nonumber\\
&+&\frac{u_2}{2}\int
d\vec r d\vec r'\, \hat\Psi^{\dag}(\vec r)\hat\Psi^{\dag}(\vec r')
{Y_{20}(\hat R)\over R^3} \hat\Psi(\vec r')\hat\Psi(\vec r),
\end{eqnarray}
with $\vec R=\vec r-\vec r'$. The Heisenberg equation
for $\hat\Psi(\vec r,t)$ becomes nonlocal.
At zero temperature the condensate wave function $\psi(\vec
r,t)=\langle\hat\Psi(\vec r,t)\rangle$ obeys the NLSE
\begin{eqnarray}
i\hbar {d\over dt}\psi(\vec r,t)
=\left[-\frac{\hbar^2}{2M}\nabla^2+V_{t}(\vec r)
-\mu+u_0|\psi(\vec r,t)|^2+u_2\int d\vec r' {Y_{20}(\hat R)\over
R^3} |\psi(\vec r',t)|^2\right]\psi(\vec r,t),
\label{nlse}
\end{eqnarray}
with $\psi(\vec r,t)$ normalized to $N$
(the number of the atom in the condensate).

\section{Numerical Studies}
In this section we discuss the ground state
properties of trapped condensates
based on numerical solutions of NLSE (\ref{nlse}).
We start with a detailed analysis of our
numerical procedure for handling the
non-local dipole interaction \cite{ml,parola,yi,goral}.

\subsection{The numerical procedure}
We use steepest descent through a propagation
of Eq. (\ref{nlse}) in imaginary time ($it$)
to find its ground state wave function.
With an axial symmetric harmonic trap
\begin{eqnarray}
V_t(\vec r)={1\over
2}M\nu^2(\lambda^2_xx^2+\lambda^2_yy^2+\lambda^2_zz^2),
\end{eqnarray}
we rescale Eq. (\ref{nlse}) by introducing dimensionless
units for length ($a_{\rm t}=\sqrt{\hbar/M\nu}$), energy
($\hbar\nu$), time ($2i/\nu$), and wave function
($\sqrt{N/a^3_{\rm t}}$). We than obtain
\begin{eqnarray}
-{d\over dt}\psi(\vec r,t)=\hat H\psi(\vec r,t),
\label{ti}
\end{eqnarray}
with
\begin{eqnarray}
\hat H=-\nabla^2+(\lambda_x^2x^2+\lambda_y^2y^2+\lambda_z^2z^2)
-2\mu+2(2\pi)^{3/2}P\left[|\psi(\vec r,t)|^2-c({\cal E})
\int d\vec r' {Y_{20}(\hat R)\over R^3} |\psi(\vec r',t)|^2\right],
\label{Hti}
\end{eqnarray}
where $P=\sqrt{2/\pi} Na_{\rm sc}/a_{\rm t}$
and $\psi(\vec r,t)$ is normalized to 1.

The ground state is found be starting with an arbitrary
random wave funtion, and propagating Eq. (\ref{ti})
in $t$ until it's stable (apart from its decreasing norm).
In practice we chose an appropriate time step
$\Delta t$ and iterates the Eq. (\ref{ti}) according to
\begin{eqnarray}
\psi(\vec r,t+\Delta t)
=\psi(\vec r,t)-(\Delta t)\hat H\psi(\vec r,t).
\label{prop}
\end{eqnarray}
We renormalize $\psi$ to 1 after each iteration
and adjust $\Delta t$ to control the rate of convergence.

For a cylindrical symmetric trap ($\lambda_x=\lambda_y=1,
\lambda_z=\lambda$), the ground state wave function
also possesses the cylindrical symmetry.
Therefore the non-local term simplifies to
\begin{eqnarray}
\int d\vec r'|\psi(\rho',z')|^2 {Y_{20}(\hat R)\over R^3}=\int
dz'\int d\rho' {\cal K}(\rho',\rho,z'-z) |\psi(\rho',z')|^2,
\end{eqnarray}
with a kernel
\begin{eqnarray}
{\cal K}(\rho',\rho,z'-z) &=&-\sqrt{5\over \pi}  {\rho' \over
{[(\rho-\rho')^2+(z'-z)^2]^2
     [(\rho+\rho')^2+(z'-z)^2 ]^{3\over 2}}}\nonumber\\
     &\ & \left(
\left[(\rho^2-\rho'^2)^2-2(\rho^2+\rho'^2)(z'-z)^2-3(z'-z)^4
\right] {\rm
E}\left[\frac{4\rho\rho'}{(\rho+\rho')^2+(z'-z)^2}\right]\right.\nonumber\\
&+&\left.[(\rho-\rho')^2+(z'-z)^2] (z'-z)^2
        {\rm K}\left[\frac{4\rho\rho'}
           {(\rho+\rho')^2+(z'-z)^2}\right] \right),\label{kn}
\end{eqnarray}
where ${\rm E}[.]$ and ${\rm K}[.]$ are standard Elliptical
integrals.

We discretize the $(\rho,z)$ plane into a two-dimensional grid of
points such that wave function values at each point becomes a
matrix. At each time step the matrix elements are updated
according to (\ref{prop}). The derivatives in the Hamiltonian are
evaluated by means of finite-difference methods. Typically, the
ground state can be sufficiently well described using a grid of
$100\times 200$ points in the range $0<\rho<5$ and $-5<z<5$.

At first sight, one may naively underestimate the
complication of Eq. (\ref{ti}) due to
the non-local interaction term in (\ref{Hti}).
Several other groups have addressed non-local
interactions previously \cite{esry,parola}.
There is however, a significant numerical
challenge with the dipole interaction, which
is singular at the origin. To accurately represent
its detailed structure, an enormously large
grid is needed. Although a Fourier transform into
momentum space could simplify the convolution operation
of the nonlocal term. We found it hard to completely
avoid the effort of the singularity this way by going to
a momentum representation with a limited coarse grid \cite{goral}.
Physically, this singularity implies the presence of two
different length scales for Eq. (\ref{ti}). We thus developed
a numerical procedure with two overlaying grids:
a coarse grid for the relatively smooth wave function
and a much finer grid for computing the non-local dipole
interaction kernel.

The kernel ${\cal K}(\rho',\rho,z'-z)$ is divergent at $\vec
r=\vec r'$, we thus define a cut-off radius $R_c$ such that ${\cal
K}(\rho',\rho,z'-z)=0$ whenever $|\vec r-\vec r'|<R_c$. This cut
off is chosen to be small enough that
negligible errors result from the numerically represented kernel.
Typically $R_c\sim 50$($a_0$) taken, which is much smaller than
the wave function grid size.
The rapid varying kernel ${\cal K}(\rho',\rho,z'-z)$ is
treated with a finer grid.
Instead of directly integrating over
${\cal K}(\rho',\rho,z'-z) |\psi(\rho',z')|^2$ on
the wave function grid, we
first integrate the kernel separately over the fine grid
around each of the wave function grid point.
Such an integration is numerically intensive,
but only needs to be performed once for each of the wave function
grid point as the kernel is determined by the geometry
of system. The integrated kernel values on the
wave function grid remain the same for different traps
and different number of atoms. Finally the non-local term
is approximated by integrating over the wave function grid
using the product of integrated kernel values
and the wave function.

For a homogeneous distribution of aligned dipoles, the mean
dipole interaction vanishes as $Y_{20}(\hat R)$ averages to zero
upon integration over $d\hat r$ or $d\hat r'$. This property is
maintained for our kernel Eq.(\ref{kn}) even though we have
integrated over $(\phi-\phi')$ first. We have verified this by
noting that the integration of ${\cal K}(\rho',\rho,z'-z)$ over
$(\rho,z)$ and $(\rho',z')$ does vanish.

\begin{figure}[h]
\centerline{\epsfig{file=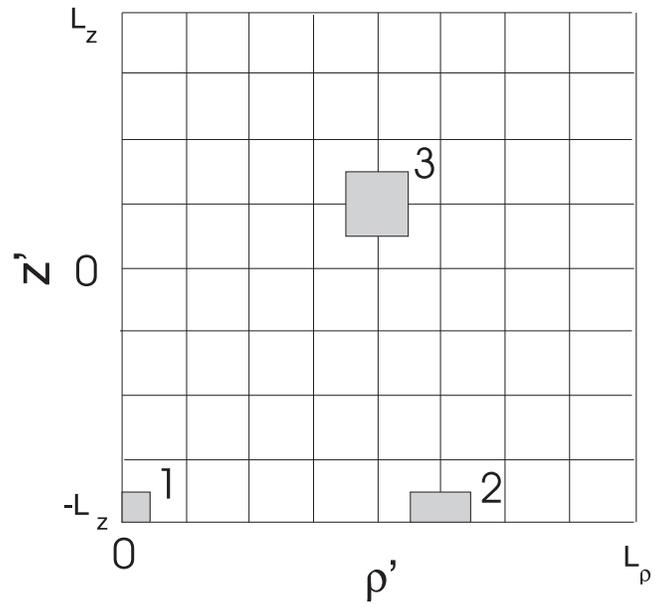,width=8.5cm}}
\caption{Computation of kernel} \label{grid}
\end{figure}

For cylindrically symmetric traps, the wave function grid as well
as the integrated kernel region is as illustrated in Fig.
\ref{grid}. An accurate representation of the integrated kernel
requires a quadrature operation over a much finer grid for each
of the shaded regions surrounding the wave function grid. As is
shown in Fig. \ref{grid}, there are three different types of
shaded regions, labeled as 1, 2, and 3. Both types of 1 and 2 are
boundary terms, which are not needed since they are respectively
at $z=\pm L_z$ and $\rho=0,L_\rho$ where either the wave function
$\psi(\rho',z')$ varnishes or the integration measure $\int \rho'
d\rho'$ vanishes. Therefore, we need only to compute the
integration of kernel over type 3 element by defining a much
finer grid and use standard numerical quadrature techniques. When
$(\rho,z)\neq(\rho',z')$, the integration reduces to a
two-dimensional one which can be easily performed. On the other
hand, a three-dimensional integration is needed when
$(\rho,z)=(\rho',z')$. In this case we have to carefully
implement the cut-off radius $R_c$.

\begin{figure}[h]
\centerline{\epsfig{file=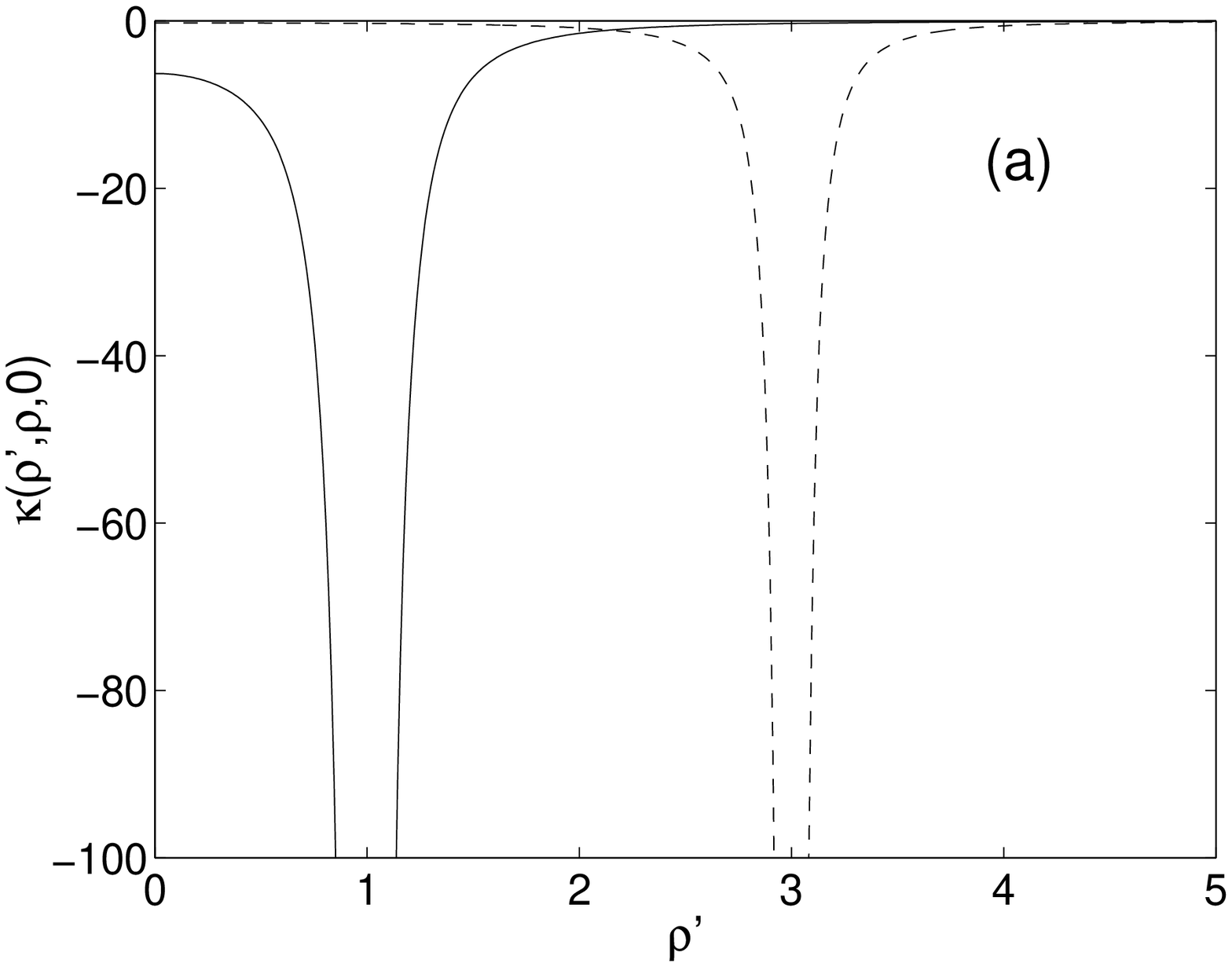,width=8.5cm}}
\centerline{\epsfig{file=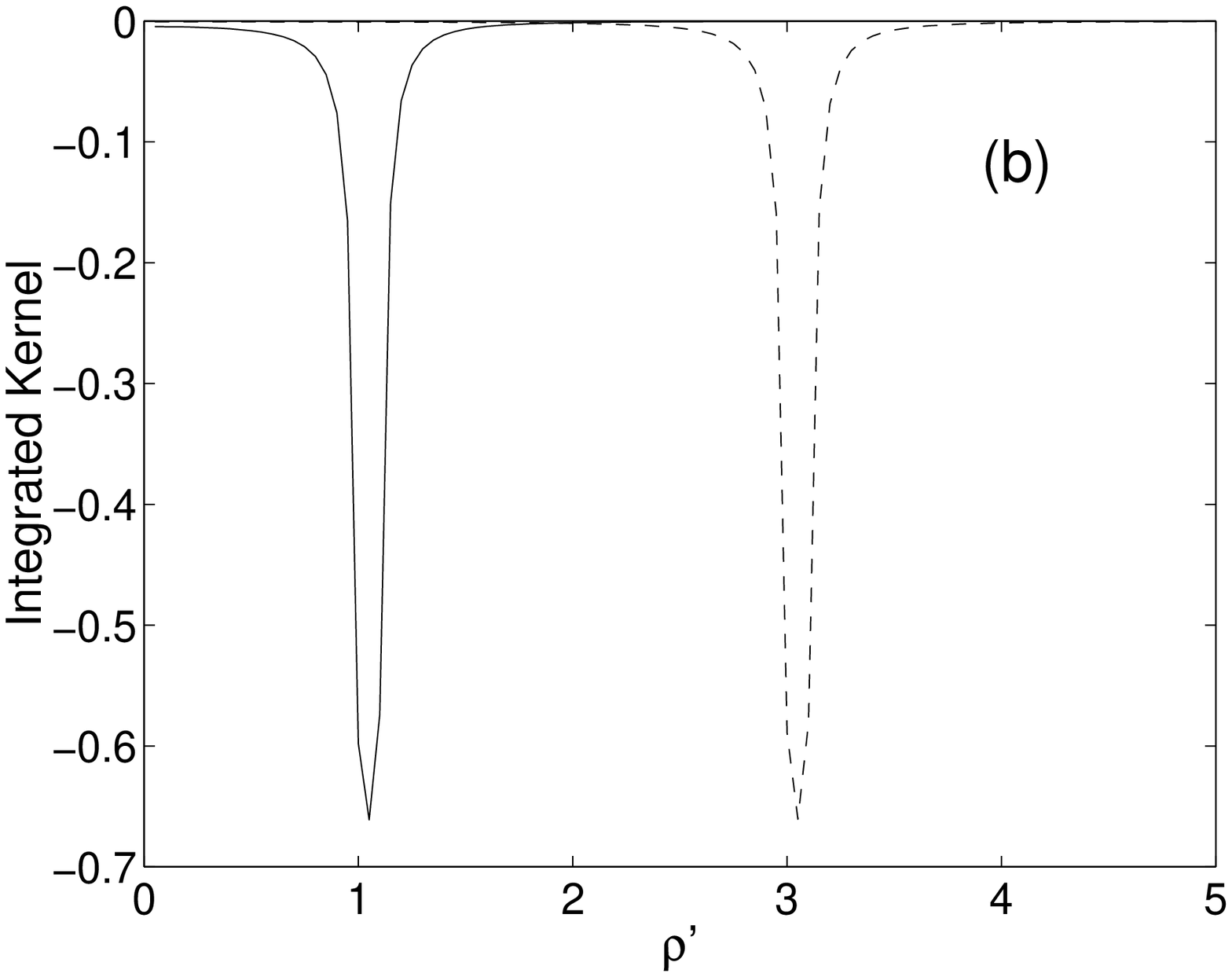,width=8.5cm}}
\caption{Typical behavior of ${\cal K}(\rho',\rho,0)$ (a) and the
corresponding integrated kernel over the wave function grid (b).
The solid (dashed) lines are for $\rho=1.0$ (3.0).}
\label{ker_int_ker}
\end{figure}

Figure \ref{ker_int_ker} compares the kernel with the
coarse grained integrated kernel. Note the significantly
different vertical scale.

\subsection{Vortex states}
Simple vortex states \cite{Cornell,Dalibard,Raman}
with quantized circulations can also be considered by
writing the condensate wave function in the form
\begin{eqnarray}
\psi(\vec r)=|\psi(\vec r)|e^{-in\phi},
\end{eqnarray}
with $\phi$ the azimuthal angle with respect to z-axis.
The corresponding Eq. (\ref{ti}) for $|\psi(\vec r)|$
is then modified by the addition of
$n^2/\rho^{2}$ to $\hat H$ of Eq. (\ref{Hti}).

\subsection{Numerical Results}
\subsubsection{Ground state wave function}
\begin{figure}[h]
\centerline{\epsfig{file=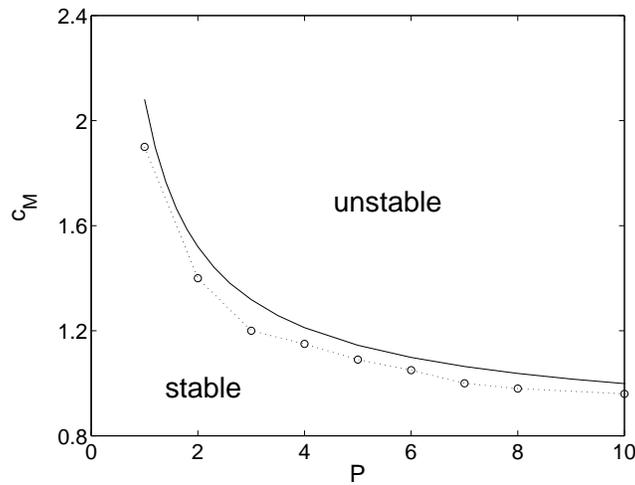,width=8.5cm}}
\caption{$c_M$ at the highest field value of condensate
collapsing for $\lambda=1$. Numerically computed values are in
circles connected by a dotted line. The solid line is from the
variational calculation.} \label{fig_cri_p_num}
\end{figure}

The ground state properties of trapped condensates with
dipole interactions were first discussed by us in \cite{yi}.
Our basic findings are: 1) condensates become
elongated along the direction of external field
while shrank in the orthogonal radial direction;
2) for given values of $P$ and $\lambda$,
 there exists a maximum $c_M(P,\lambda)$ of
 a threshold field strength beyond which condensate collapse
 occurs. Figure \ref{fig_cri_p_num} displays
numerically computed $c_M$ for $\lambda=1$ at several
different $P$ values. For comparison we
also show the results from variational
calculations to be discussed later.

The condensate collapse is mainly due to attractive dipole
interactions along the direction of external field. To minimize
its total energy, condensed atoms prefer to align along
the attractive direction (z-axis), while narrowing
its width along the radially repulsive direction.
The collapsing occurs when the radial width eventually
approaches zero with increasing external field strength \cite{pit}.

\begin{figure}[h]
\centerline{\epsfig{file=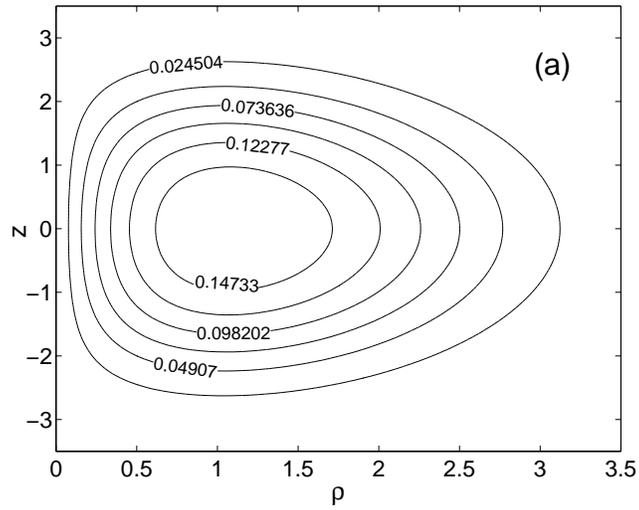,width=8.5cm}}
\centerline{\epsfig{file=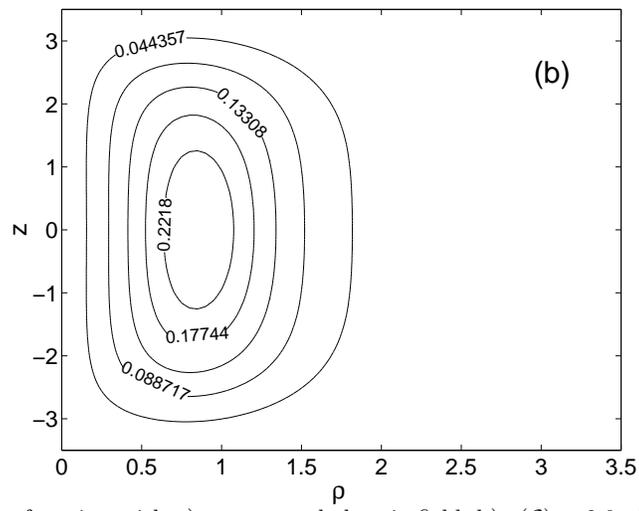,width=8.5cm}}
\caption{$n=1$ vortex state wave function with a) no external
electric field; b)
 $c({\cal E})=2.0$. Other parameters are
$P=10$ and $\lambda=1$.} \label{fig_vortex}
\end{figure}

Figure \ref{fig_vortex} shows a vortex state wave function for
$n=1$. The effects of dipole interactions are similar to the
ground state. As $c({\cal E})$ increases, the vortex state will
also collapse. Because of the zero density inside the vortex core,
$c_M$ in this case is much larger than for the ground state.

\subsubsection{Comparing numerical solutions with TFA}

A useful approximation for the ground state solution
of NLSE (\ref{nlse}) is the Thomas-Fermi approximation
(TFA). When the interaction between atoms are repulsive, {\it
i.e.} with a positive scattering length, the condensate
is expected to increase its size as compared
to the single atom ground state in the trap.
With more atoms, the larger the condensate size,
eventually the spatial directives, consequently the kinetic
energy term becomes negligible.
In this limit, TFA is used to find the ground state wave function
by neglecting the kinetic energy term in Eq. (\ref{Hti}).
With a SIA interaction term, the solution simply
takes the shape of the inverted trap potential $V_t(\vec r)$.
The nonlocal dipole interaction term, however, prevents a simple
analytic solution even with the TFA.

\begin{figure}[h]
\centerline{\epsfig{file=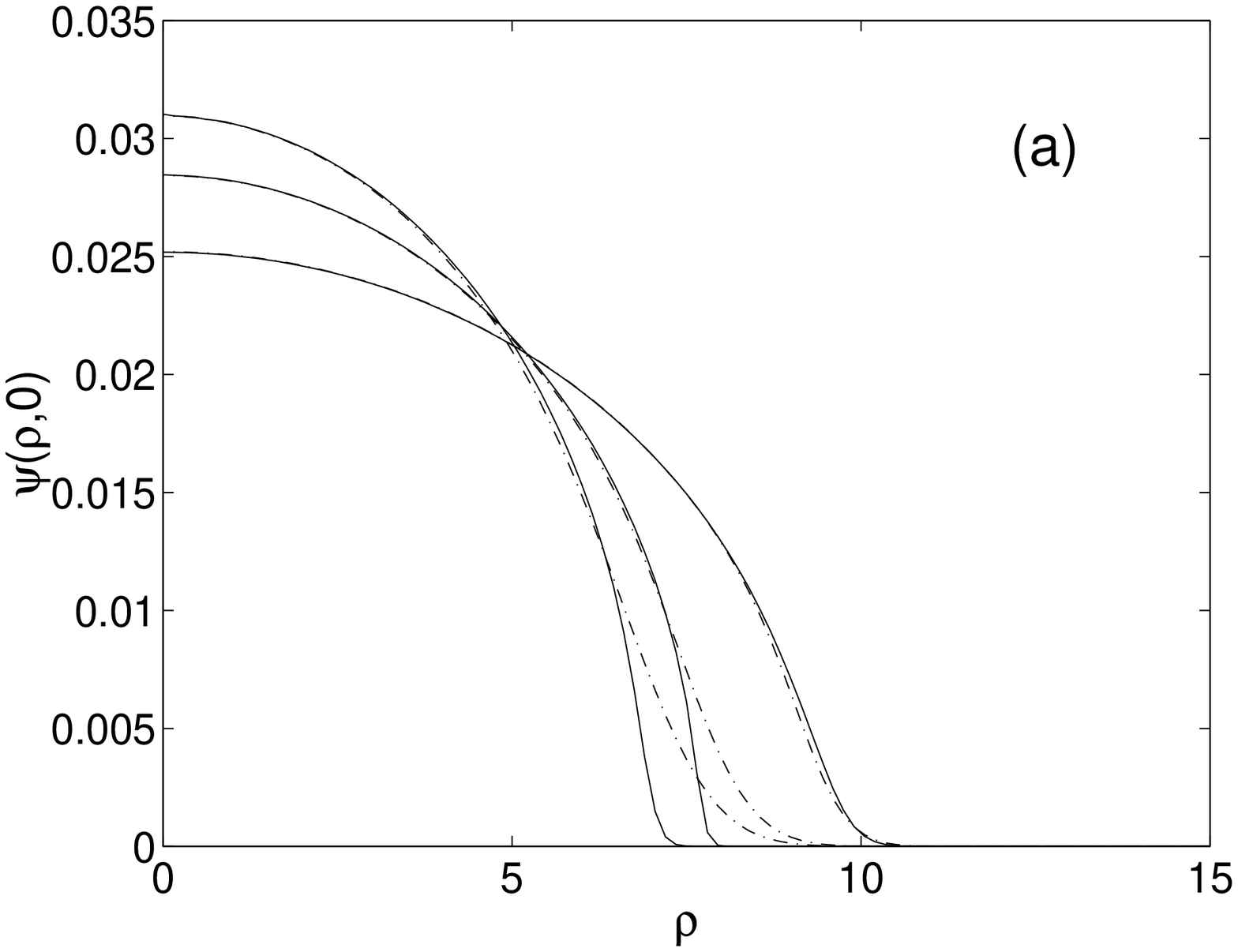,width=8.5cm}}
\centerline{\epsfig{file=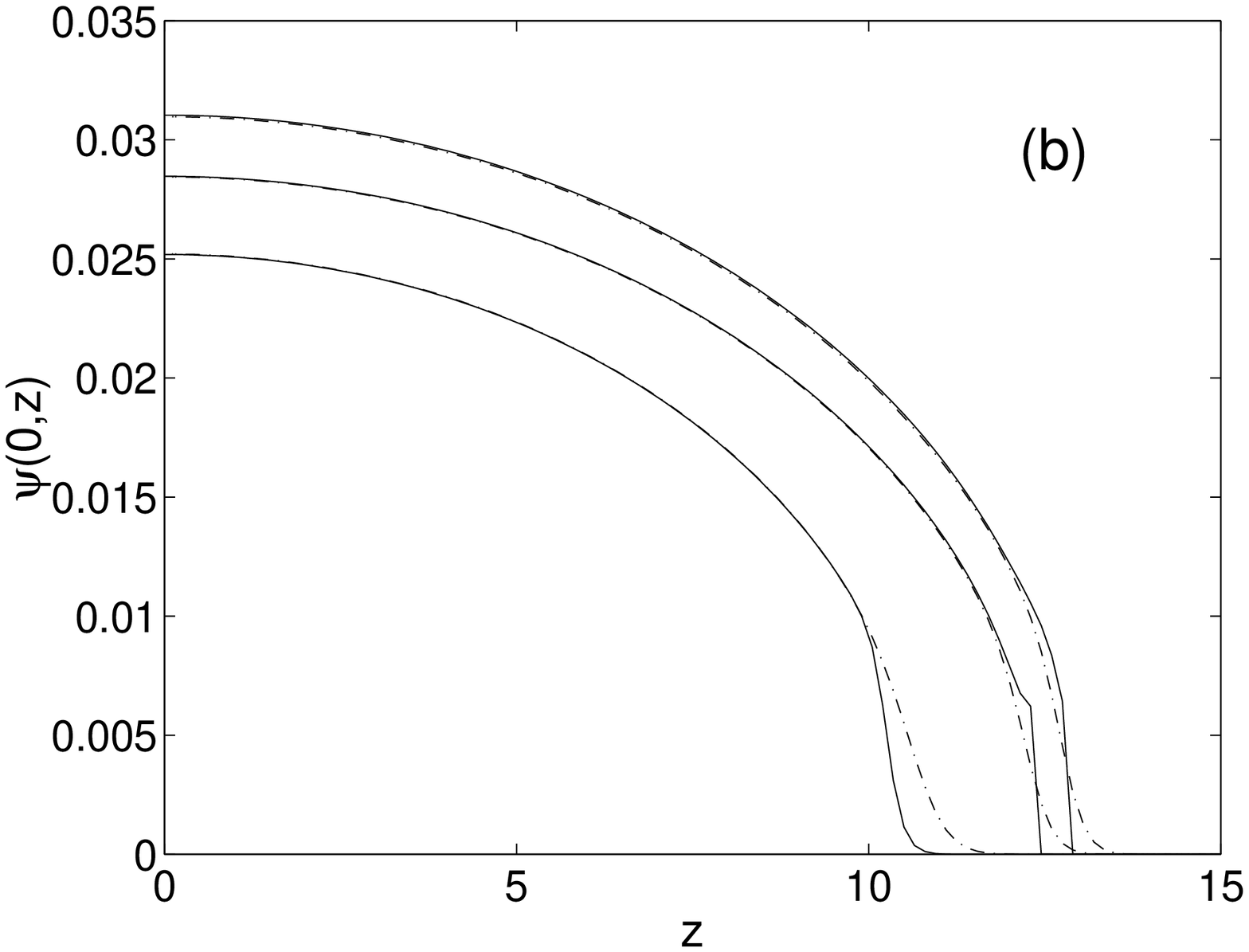,width=8.5cm}}
\caption{Ground state wave functions with (solid line) and
without (dash-dotted line) TFA. $P=5000$, $\lambda=1$ and
$c({\cal E})=0.7,0.6,0.2$ in descending order of wave function
values at the origin.} \label{tfa_num}
\end{figure}

Typical solutions to Eq. (\ref{nlse}) with and without TFA
for $P=5000$, $\lambda=1$, and $c({\cal E})=0.2,0.6,0.7$
are compared in Fig. \ref{tfa_num}.

\section{time-dependent variational analysis}
The time-dependent variational approach can also be
used to analyze solutions of (\ref{nlse}) \cite{var,var1}.
We start by identifying a Lagrangian density ${\cal L}$
\begin{eqnarray}
{\cal L}&=&{i\over 2}\hbar\left[\psi(\vec r){\partial\psi^*(\vec
r)\over\partial t}-\psi^*(\vec r){\partial\psi(\vec
r)\over\partial t}\right]\nonumber\\
&+&{\hbar^2\over 2M}|\nabla\psi(\vec
r)|^2+V_{\rm t}(\vec r)|\psi(\vec r)|^2\nonumber\\
&+&{u_0\over 2}|\psi(\vec r)|^4+{u_2\over 2}|\psi(\vec r)|^2\int d{\vec
r'}{Y_{20}(\hat R)\over R^3}|\psi(\vec r')|^2.\label{lagden}
\end{eqnarray}
The NLSE can then be found from a minimization of the
action \cite{var1}
\begin{eqnarray}
S=\int{\cal L}d{\vec r}dt.
\end{eqnarray}
To simplify the variational calculation,
we restrict $\psi$ to a convenient family of trial functions and
study the time evolution of the parameters that define the
family. A natural choice is a Gaussian-like function first
used in \cite{var1}
\begin{eqnarray}
\psi(x,y,z,t)=A(t)\prod_{\eta=x,y,z}e^{-[\eta
-\eta_0(t)]^2/2w^2_\eta+i\eta\alpha_\eta(t)+i\eta^2\beta_\eta(t)},\label{gausan}
\end{eqnarray}
where $A$ (complex amplitude), $w_\eta$ (width), $\alpha_\eta$
(slope), $\beta_\eta$ (curvature radius)$^{-1/2}$, and $\eta_0$
(center of cloud) are variational parameters. This approach,
pioneered by Perez-Garcia {\it et al.} \cite{var1}, has since
been successfully used for many studies of trapped condensates,
a more recent application attempted to explain the
anomalous behavior in the finite temperature excitation
experiment \cite{henk,jin}.

Our goal here is to find equations governing the variational parameters.
To this aim, we insert (\ref{gausan}) into (\ref{lagden}) and
calculate an effective Lagrangian $L$ by integrating the Lagrangian
density over all space coordinates
\begin{eqnarray}
L=\langle{\cal L}\rangle=\int^\infty_{-\infty}{\cal L}d{\vec r},
\end{eqnarray}
to arrive at
\begin{eqnarray}
L&&={i\hbar N\over 2}\left({\dot{A^*}\over A^*}-{\dot{A}\over
A}\right)\nonumber\\&&+{N\over
2}\sum_\eta\left[\left(\hbar\dot{\beta_\eta}+{2\hbar^2\beta^2_\eta\over
M}+{1\over
2}M\nu^2\lambda^2_\eta\right)(w^2_\eta+2\eta^2_0)+\left(\hbar\dot{\alpha_\eta}
+{2\hbar^2\alpha_\eta\beta_\eta\over M}\right)2\eta+{\hbar^2\over
2Mw^2_\eta}+{\hbar^2\alpha^2_\eta\over
M}\right]\nonumber\\&&+{N^2\over
2\sqrt{8}\pi^{3/2}w_xw_yw_z}\left[u_0+u_2\sqrt{5\over 16\pi}\int
d{\vec r}{\exp}\left\{-\sum_\eta{\eta^2\over
2w^2_\eta}\right\}{3{\cos}^2\theta-1\over r^3}\right],
\end{eqnarray}
where we have used atom number conservation
\begin{eqnarray}
N=\pi^{3/2}|A(t)|^2w_x(t)w_y(t)w_z(t)={\rm const}.
\end{eqnarray}

At this point, the variation calculation
effectively has been reduced to a finite dimensional
problem, i.e., to solve the Lagrange equations
\begin{eqnarray}
{d\over dt}\left({\partial L\over \partial
\dot{q_j}}\right)-{\partial L\over
\partial q_j}=0,
\end{eqnarray}
with the notation
\begin{eqnarray}
q_j\equiv\{w_x,w_y,w_z,A,A^*,x_0,y_0,z_0,\alpha_x,\alpha_y,\alpha_z,\beta_x,\beta_y,\beta_z\}.
\end{eqnarray}
We find equations for
the center of the condensate
\begin{eqnarray}
\ddot{\eta_0}+\lambda^2_\eta\nu^2\eta_0=0,
\end{eqnarray}
and the condensate widths satisfy
\begin{eqnarray}
\ddot{w_\eta}+\lambda^2_\eta\nu^2w_\eta ={\hbar^2\over
M^2w^3_\eta}-{N\over 4\sqrt{2}\pi^{3/2}M}{\partial\over
\partial w_\eta}\left[{1\over w_xw_yw_z}\left(u_0+u_2\sqrt{5\over
16\pi}\int d{\vec r}{\exp}\left\{-\sum_\eta{\eta^2\over
2w^2_\eta}\right\}{3{\cos}^2\theta-1\over r^3}\right)\right].
\end{eqnarray}
The rest of the parameters can be obtained from
\begin{eqnarray}
\beta_\eta={M\dot{w_\eta}\over 2\hbar w_\eta},
\end{eqnarray}
and
\begin{eqnarray}
\alpha_\eta={M\over
\hbar}\left(\dot{\eta_0}-{\eta_0\dot{w_\eta}\over w_\eta}\right).
\end{eqnarray}

It is convenient to introduce new dimensionless variables
$\tau=\nu t$ and $w_\eta=a_{\rm t} v_\eta$. We then arrive at
\begin{eqnarray}
{d^2\over d\tau^2}v_\eta+\lambda^2_\eta v_\eta={1\over
v^3_\eta}-P{\partial\over\partial v_\eta}\left[{1\over
v_xv_yv_z}\left(1-\sqrt{5\over 16\pi}c({\cal E})\int d{\vec r}{\rm
exp}\left\{-\sum_\eta{\eta^2\over 2v^2_\eta}\right\}{3{\rm
cos}^2\theta-1\over r^3} \right)\right],\label{width}
\end{eqnarray}
where $P=\sqrt{2/\pi}Na_{\rm sc}/a_{\rm t}$. This equation
describes the motion of a particle with coordinates
$(v_x,v_y,v_z)$ in an effective potential
\begin{eqnarray}
U(v_x,v_y,v_z)&&={1\over
2}\left(\lambda^2_xv^2_x+\lambda^2_yv^2_y+\lambda^2_zv^2_z\right)+{1\over
2}\left({1\over v^2_x}+{1\over v^2_y}+{1\over
v^2_z}\right)\nonumber\\&&+{P\over v_xv_yv_z}\left[1-\sqrt{5\over
16\pi}c({\cal E})\int d{\vec r}{\exp}
\left\{-\sum_\eta{\eta^2\over 2v^2_\eta}\right\}
{3{\cos}^2\theta-1\over r^3} \right].
\label{eqv}
\end{eqnarray}

For a cylindrically symmetric trap with $\lambda_x=\lambda_y=1$,
$\lambda_z=\lambda$, all integrals can be performed analytically
to yield
\begin{eqnarray}
{d^2\over d\tau^2}v+v&&={1\over v^3}+{P\over v^3v_z}\left[1-c({\cal
E})f(\kappa)\right],\\ \label{evov1}
{d^2\over d\tau^2}v_z+\lambda^2v_z&&={1\over v^3_z}+{P\over
v^2v^2_z}\left[1-c({\cal E})g(\kappa)\right],\label{evovz1}
\end{eqnarray}
with $v_x=v_y=v$, $\kappa=v/v_z$, and
\begin{eqnarray}
f(\kappa) &&={\sqrt{5\pi}\over 6(1-\kappa^2)^2}
\left[-4\kappa^4-7\kappa^2+2+9\kappa^4H(\kappa)\right],\nonumber\\
g(\kappa) &&={\sqrt{5\pi}\over 3(1-\kappa^2)^2}
\left[-2\kappa^4+10\kappa^2+1-9\kappa^2H(\kappa)\right].
\end{eqnarray}
$H(\kappa)\equiv{{\tanh}^{-1}\sqrt{1-\kappa^2}/
\sqrt{1-\kappa^2}}$.
The equilibrium widths are then determined by
\begin{eqnarray}
v_0 &&={1\over v^3_0}+{P\over v^3_0v_{z0}}
\left[1-c({\cal E})f(\kappa_0)\right], \nonumber\\
\lambda^2v_{z0} &&={1\over v^3_{z0}}+{P\over
v^2_0v^2_{z0}}\left[1-c({\cal E})g(\kappa_0)\right].
\label{ev1}
\end{eqnarray}
Using $v_{z0}$ and $\kappa_0$, Eq. (\ref{ev1}) can be
rewritten as
\begin{eqnarray}
\kappa_0 v_{z0} &&={1\over \kappa_0^3 v^3_{z0}}+{P\over
\kappa_0^3v^4_{z0}}\left[1-c({\cal E})f(\kappa_0)\right], \nonumber\\
\lambda^2 v_{z0} &&={1\over v^3_{z0}}+{P\over
\kappa_0^2v^4_{z0}}\left[1-c({\cal E})g(\kappa_0)\right].
\label{eqrootkv}
\end{eqnarray}

In following discussion, we consider only $a_{\rm sc}({\cal E})
>0$ case, which implies both $P>0$ and $c({\cal E}) >0$.

\subsection{Equilibrium widths} 
\begin{figure}[h]
\centerline{\epsfig{file=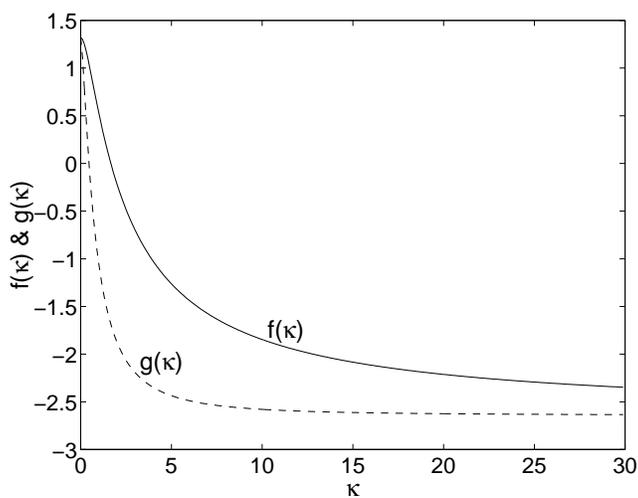,width=8.5cm}}
\caption{$f(\kappa)$,$g(\kappa)$} \label{fg}
\end{figure}

\begin{figure}[h]
\centerline{\epsfig{file=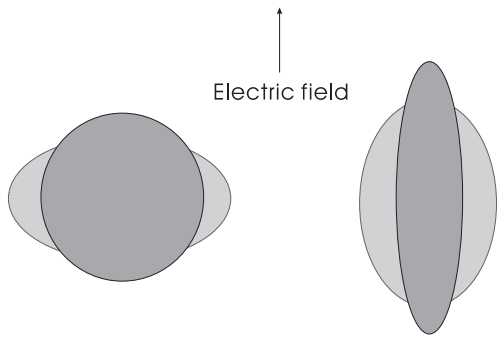,width=6.5cm}\\[9pt]}
\caption{Condensates (pancake shaped in the left and cigar shaped
to the right) always expands along the direction of an externally
applied electric field and shrinks along the repulsive radial
directions. Darker ellipses are for higher electric fields.}
\label{shape}
\end{figure}

First, for simplicity, we assume
that our system satisfy Thomas-Fermi limit ($P\gg 1$), then we
can safely ignore the kinetic term and rewrite Eq.
(\ref{eqrootkv}) for $\kappa_0$ in the following form
\begin{eqnarray}
\kappa_0^2\left[1-c({\cal
E})g(\kappa_0)\right]=\lambda^2\left[1-c({\cal
E})f(\kappa_0)\right].\label{rootk}
\end{eqnarray}
This equation can be solved graphically. From Fig. \ref{fg}, we
first note both $f(\kappa)$ and $g(\kappa)$ are monotonically
decreasing functions bounded between $\sqrt{5\pi}/3$ and
$-2\sqrt{5\pi}/3$, also the inequality $f(\kappa)>g(\kappa)$ holds
for all $\kappa>0$. Then for all $\kappa\geq\lambda$, we have
$\kappa^2\left[1-c({\cal E})
g(\kappa)\right]>\lambda^2\left[1-c({\cal E})f(\kappa)\right]$,
therefore, if $\kappa_0$ is a solution of Eq. (\ref{rootk}), it
must satisfy $\kappa_0<\lambda$. Meanwhile, when $c({\cal E})=0$,
the solution for $\kappa$ in Eq. (\ref{rootk}) is
$\kappa_0=\lambda$. This then proves that no matter what the
initial field-free condensate aspect ratio is, the condensate
always become more prolate along the electric field direction,
{\it i.e.} approximately as illustrated in Fig. \ref{shape}, it
expands along the field direction but shrinks in the orthogonal
direction. As will be discussed in more detail later, the total
condensate volume actually shrinks with increasing fields because
of the attractive dipole interaction.

This result, first explained by us \cite{yi} in terms of the
minimization of total energy, is different from the conclusion
reached in Ref. \cite{goral} based on a force argument. This
interesting feature has also been independently verified by
numerical solutions based on the FFT algorithm adopted by Goral
{\it et. al} \cite{goral}.

\begin{figure}[h]
\centerline{\epsfig{file=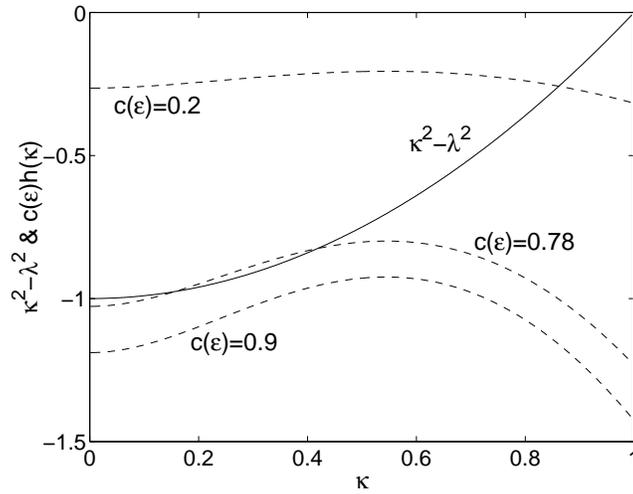,width=8.5cm}} \caption{Graphic
solutions of Eq. (\ref{eqrootk2}) for $\lambda=1$. Functions
$\kappa^2-\lambda^2$ (solid line) and $c({\cal E})h(\kappa)$
(dashed line) for different values of $c({\cal E})$.}
\label{figce}
\end{figure}

We can also rewrite Eq. (\ref{rootk}) as
\begin{eqnarray}
\kappa^2-\lambda^2=c({\cal E})h(\kappa),\label{eqrootk2}
\end{eqnarray}
with $h(\kappa)=\kappa^2g(\kappa)-\lambda^2f(\kappa)$. Figure
\ref{figce} shows its graphical solutions ($\lambda=1$) at
several different $c({\cal E})$ values.

First, we note that $h(\kappa=0)=-\sqrt{5\pi}\lambda^2/3$. Thus,
as long as $c({\cal E})<3/\sqrt{5\pi}$, there will be one and
only one root. This result may not look absolutely clear from the
figure because of plotting constraints, but it can be seen
clearly form Eq. (\ref{rootk}). We also find that as $c({\cal
E})$ increases, there may exist one, two, or zero roots for
$\kappa$. Once $\kappa_0$ is known, one can easily find solutions
$v_0$ and $v_{z0}$, whose stability can also be checked
straightforwardly. It turns out that, in all our calculations,
whenever only one root for $\kappa$ occurs its corresponding
solution for $v_0$ and $v_{z0}$ is always stable. If there are
two roots for $\kappa$, the solution for $v_0$ and $v_{z0}$
corresponding to the smaller $\kappa_0$ root is a saddle point,
thus always unstable, while the other is always stable.

\begin{figure}[h]
\centerline{\epsfig{file=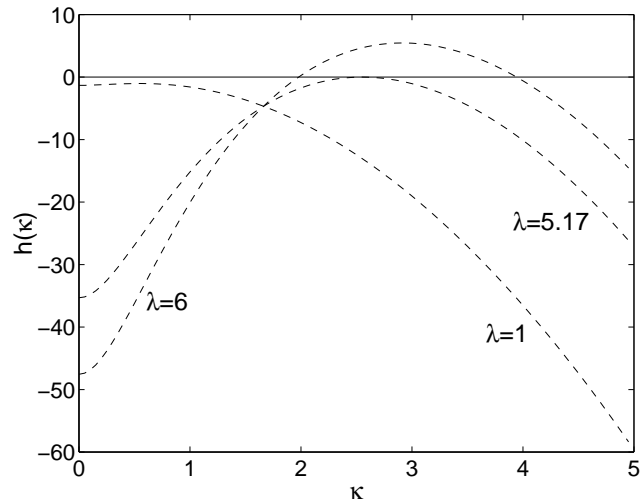,width=8.5cm}} \caption{Function
$h(\kappa)$ for different $\lambda$ value.} \label{figlamh}
\end{figure}

\begin{figure}[h]
\centerline{\epsfig{file=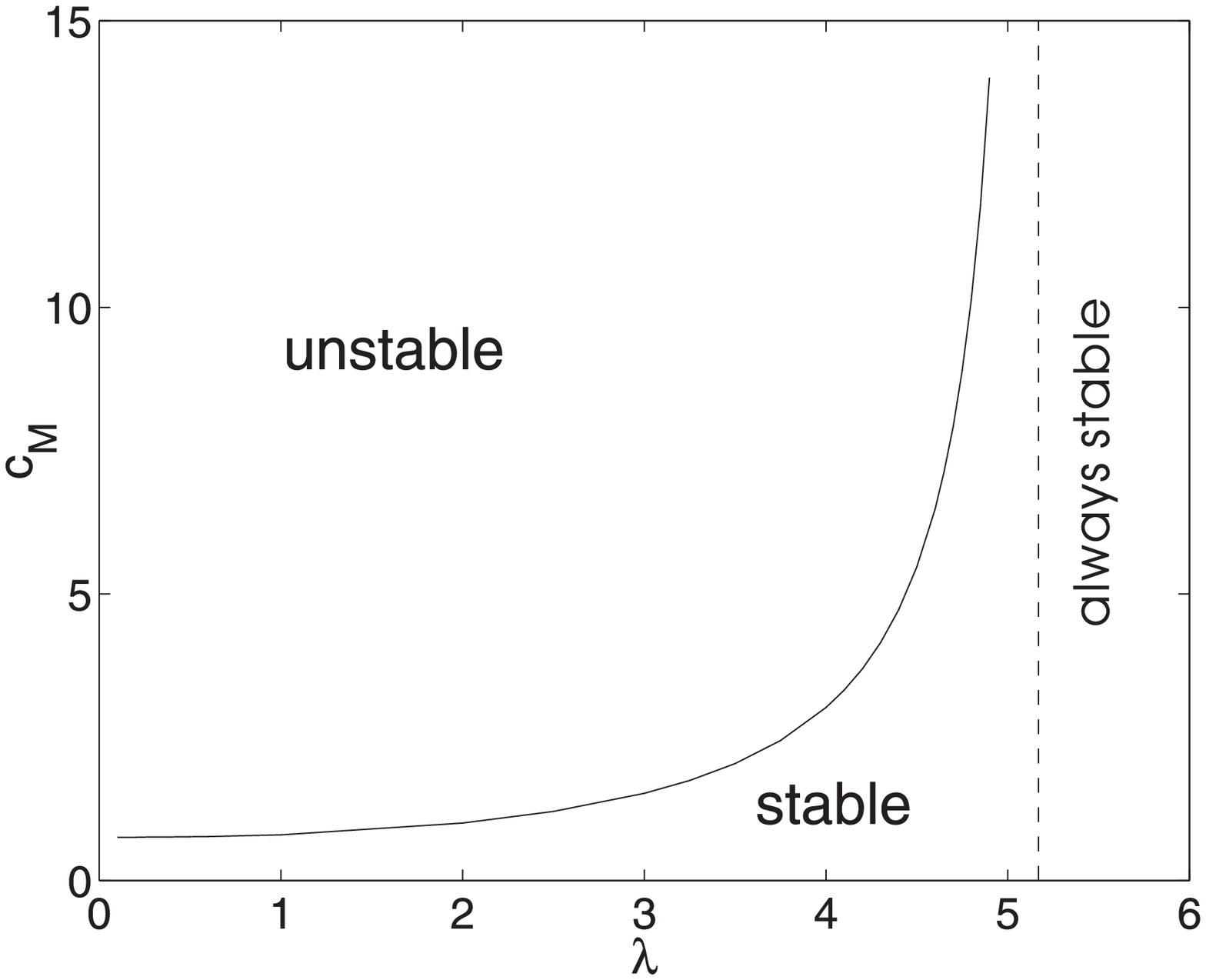,width=8.5cm}}
\caption{$c_M$ as a function of $\lambda$ with TFA.}
\label{cri_ce_la}
\end{figure}

We now consider the $\lambda$ dependence of the condensate
property. Figure \ref{figlamh} shows the function $h(\kappa)$ at
several different $\lambda$ values. We see that as $\lambda$
increases, the maximum value of function $h(\kappa)$, $h_{\rm
max}$ also increases, and $h_{\rm max}=0$ at $\lambda_c=5.170169$.
For those $\lambda$ values corresponding to a negative $h_{\rm
max}$, no root of $\kappa$ is found if $c({\cal E})$ becomes
sufficiently large. But when $\lambda\geq\lambda_c$, at least one
root for $\kappa$ exists no matter how large a $c({\cal E})$ is.
This means the condensate cannot simply collapse even at these
very large electric field strengths. Physically this implies the
increased stability of condensate inside an electric field with
increasing values of $\lambda$. A condensate of a pancake shape
is more stable than one with a cigar shape. This can be simply
understood from the following argument; The collapse of a
condensate under electric fields is due to the alignment of atoms
along the attractive z-axis direction. A larger $\lambda$ value
prevents such alignment which increases both kinetic as well as
trap potential energy, hence increases the stability. In Fig.
\ref{cri_ce_la}, we display $c_M$ as a function of $\lambda$,
which separates the stable and unstable regions. Rigorous
numerically calculations, on the other hand, find that condensate
can still collapse even when $\lambda\geq\lambda_c$. For example,
we found collapse occurs when $P\ge 5000$ with $\lambda>7$,
[$c({\cal E})>2.0$]. This difference maybe due to the choice of a
simple Gaussian shaped variation function (\ref{gausan}).

\begin{figure}[h]
\centerline{\epsfig{file=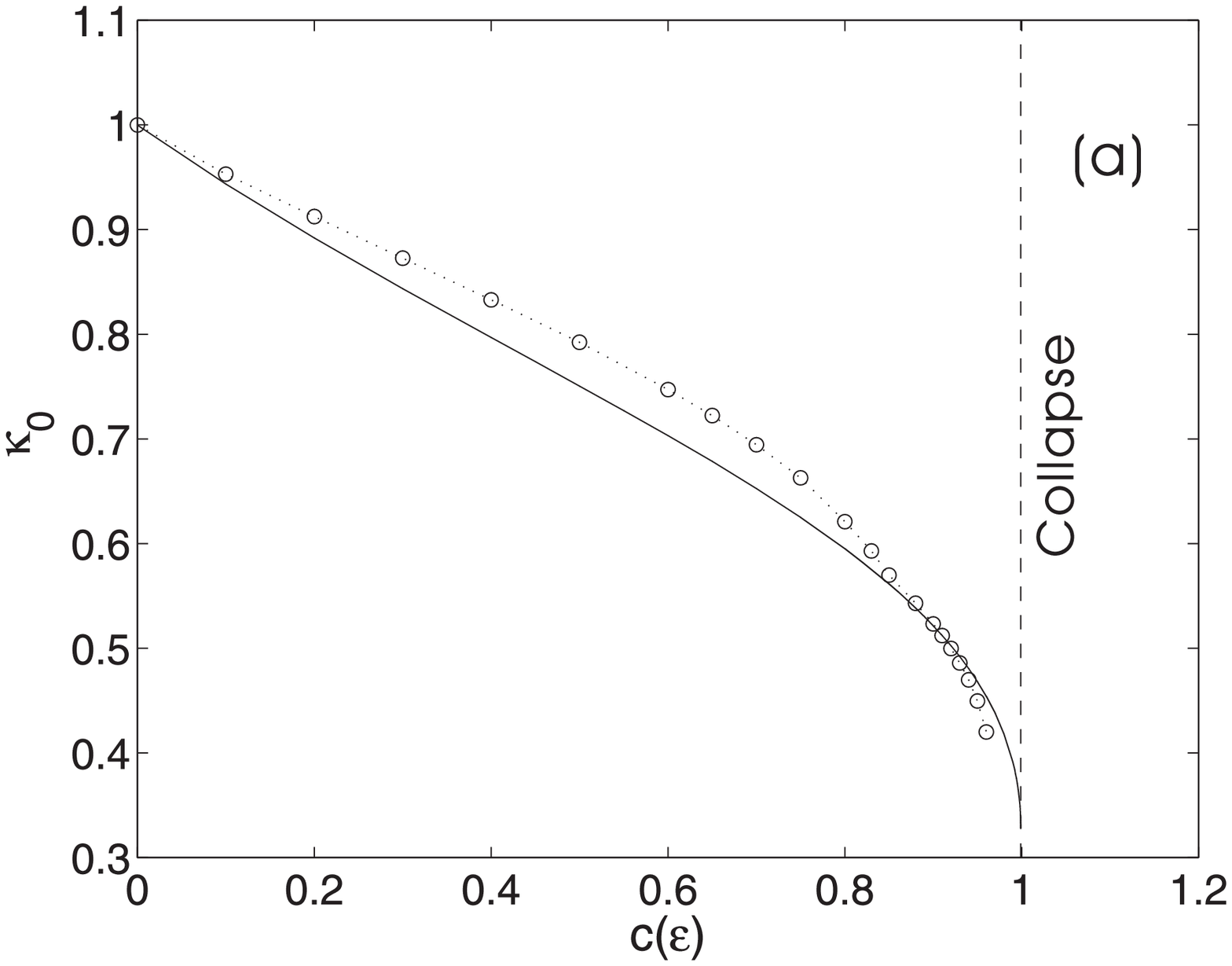,width=8.5cm}}
\centerline{\epsfig{file=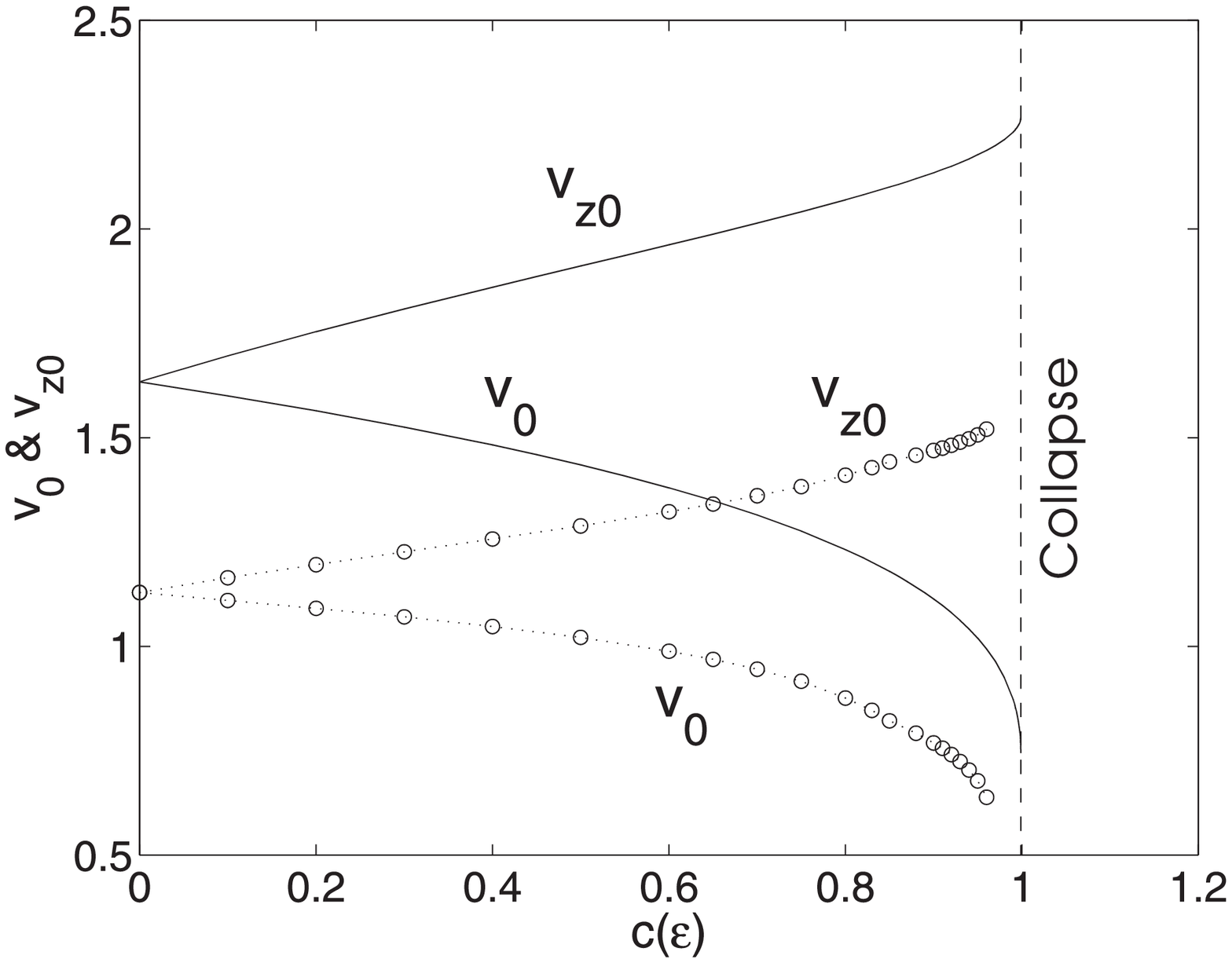,width=8.5cm}}
\centerline{\epsfig{file=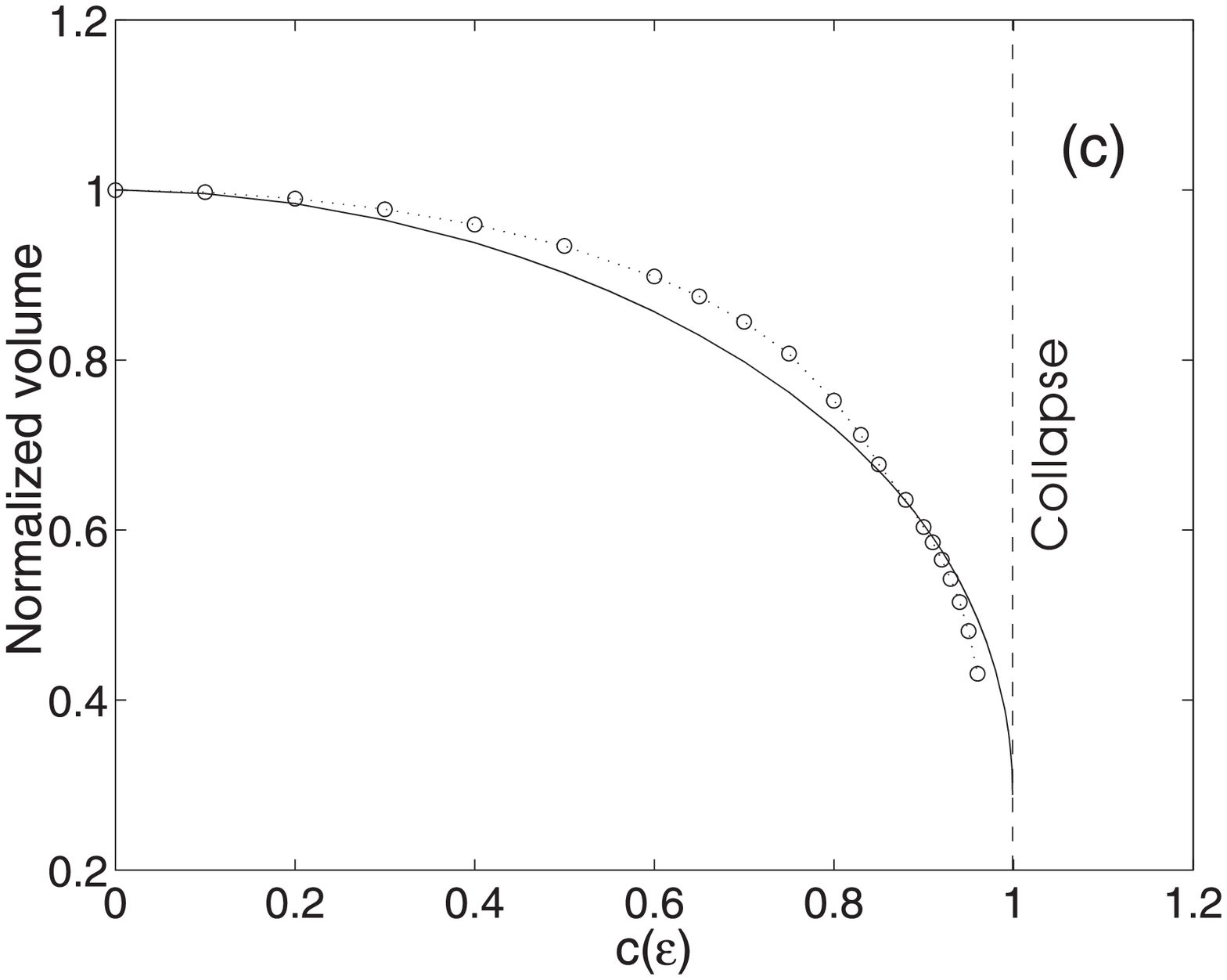,width=8.5cm}} \caption{Field
dependence of $\kappa_0$ (a), $v_0$, $v_{z0}$ (b) and normalize
volume (c) of condensate for $P=10$, $\lambda=1$. Numerically
computed results are in circles connected by dotted lines, the
solid lines are from variational calculations. }
\label{fig_widths}
\end{figure}

When the system is not in the Thomas-Fermi limit, we have to first
solve for $\kappa_0$ from equation
\begin{eqnarray}
\kappa_0^2-\lambda^2=c({\cal E})
[\kappa_0^2g(\kappa_0)-\lambda^2f(\kappa_0)]+\left\{{(\kappa_0^4-\lambda_0^2)^4\over
P^4\kappa_0^2 }\left[1-\kappa_0^2-c({\cal E})
(g(\kappa_0)-\kappa_0^2f(\kappa_0))\right]\right\}^{1/5},
\end{eqnarray}
and then using $$ v_{z0}=P{\kappa_0^2\left[1-c({\cal
E})g(\kappa_0)\right]-\lambda^2\left[1-c({\cal
E})f(\kappa_0)\right]\over \lambda^2-\kappa_0^4},$$ and
$v_0=\kappa_0 v_{z0}$ to find the equilibrium widths. In this
case, we find possibilities for one, two, three, and four or no
roots of $\kappa_0$ depending on values of $P$, $c({\cal E})$,
and $\lambda$. The stability conditions for these roots, however,
are similar to the TFA case as discussed before---there is at most
only one stable solution. In Fig. \ref{fig_widths}, we present
$\kappa_0$, $v_0$, and $v_{z0}$ as functions of $c({\cal E})$. We
see that $v_0$ decreases with increasing $c({\cal E})$, and the
condensate collapses when $v_0$ goes to zero. Figure
\ref{fig_widths}(c) displays the dependence of the condensate
volume on ${\cal E}$, where the volume is defined as the produce
of its effective widths in three separate dimensions. In terms of
absolute values, numerically calculated volume (averaged width)
 differs from the variational result by a few times due to
the multiplying effects of three widths.
The shrinking volume increases the condensate density,
which in turn can significantly increase the
three body loss process, providing a potentially
useful diagnosis tool \cite{frec}.

\begin{figure}[h]
\centerline{\epsfig{file=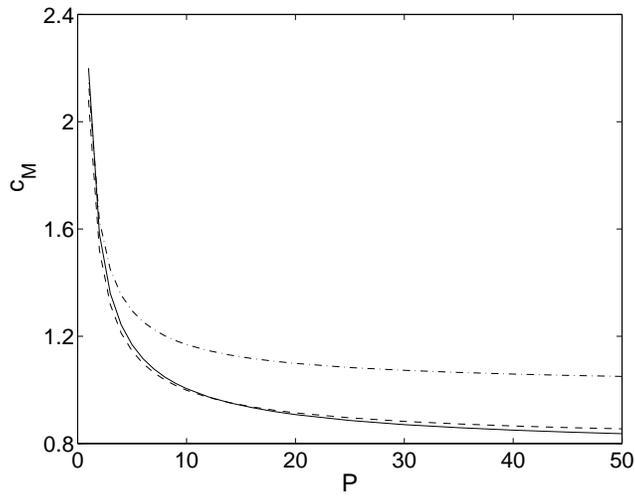,width=8.5cm}}
\caption{$c_M$ as a function of $P$, for $\lambda=0.5$ (solid),
$1.0$ (dashed), and $2.0$ (dash-dotted line). As
$P\rightarrow\infty$, $c_M$ goes to the value obtained under
TFA.} \label{cri_ce_p}
\end{figure}

Figure \ref{cri_ce_p} shows $c_M$ as a function of $P$ for
different $\lambda$ values. We see that, for smaller $P$, a
condensate with a small $\lambda$ can be more stable than
condensates with larger $\lambda$; while for larger $P$, a
condensate with a larger $\lambda$ is always more stable than
condensates with smaller $\lambda$. This also confirmed by both
variational (TFA) and numerical calculations.

\begin{figure}[h]
\centerline{\epsfig{file=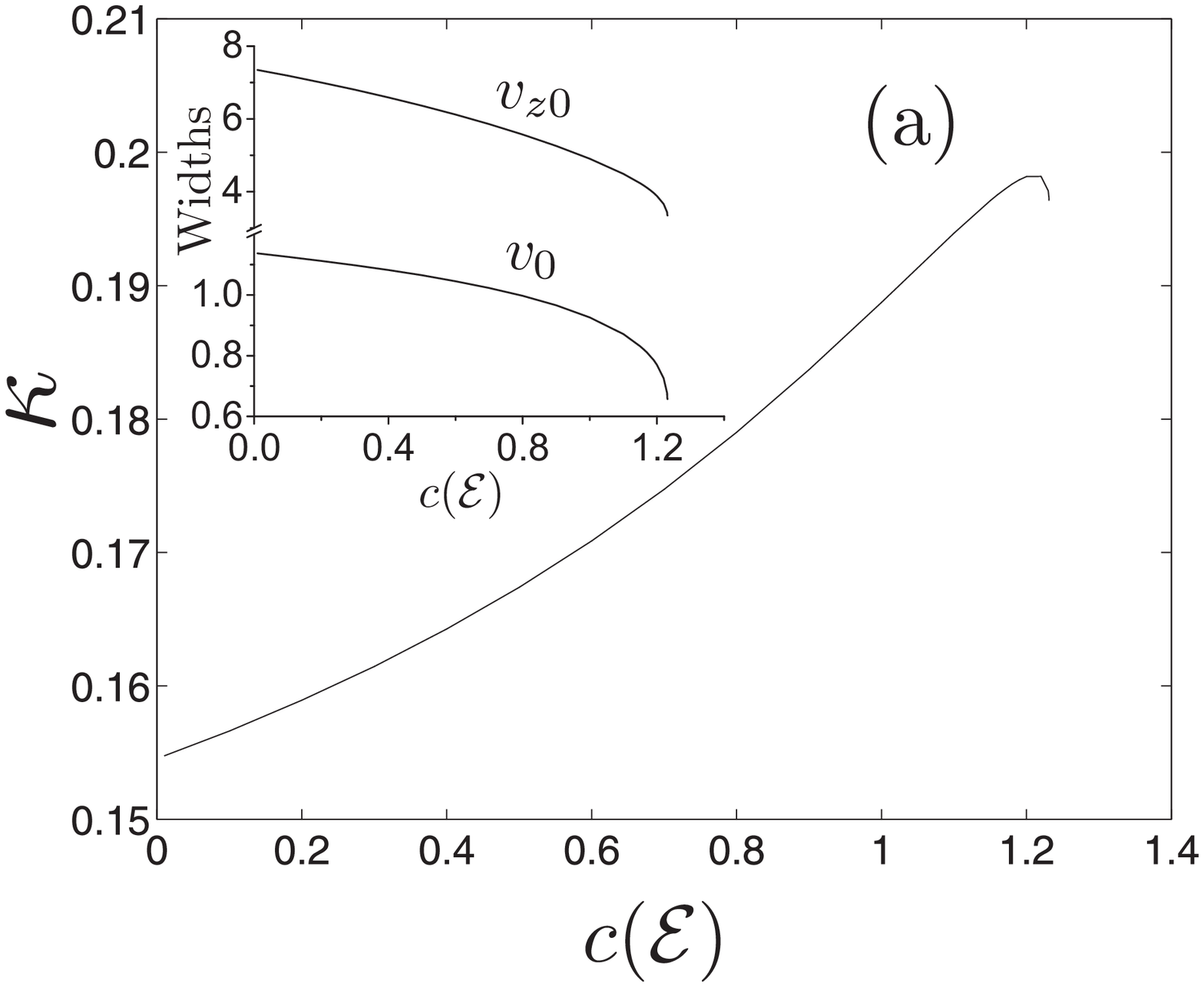,width=8.5cm}}
\centerline{\epsfig{file=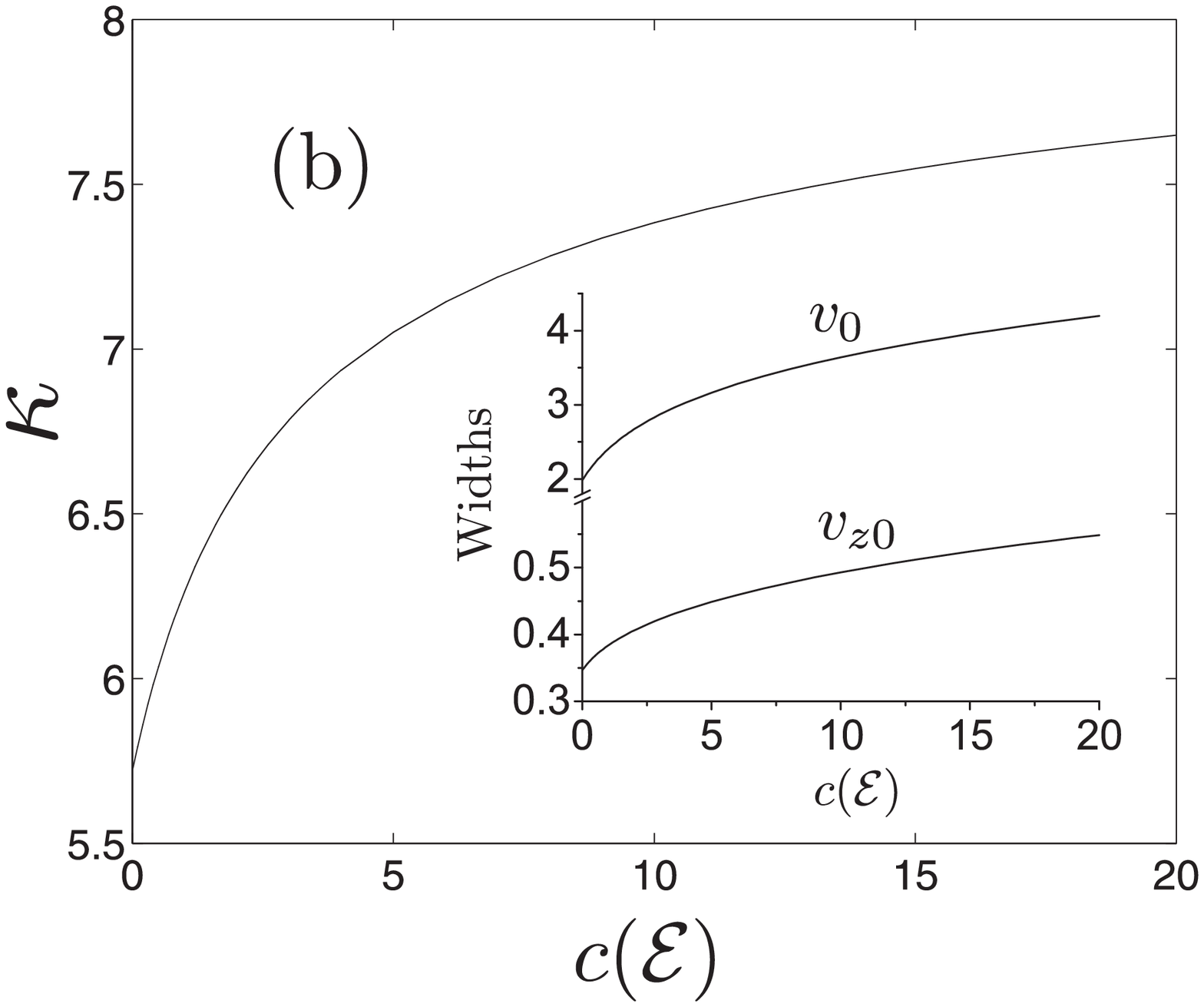,width=8.5cm}}
\caption{Field dependence of $\kappa_0$ and condensate widths (in
insets) for $P=5$, $\lambda=0.1$ (a) and $\lambda=10$ (b). When
$\lambda<\lambda_c$, the volume of condensate decreases with
$c(\cal E)$. It increases with $c(\cal E)$ when
$\lambda>\lambda_c$, causing condensate to be always stable.}
\label{kappa_p5}
\end{figure}

After we first submitted this paper, a paper on the same topic
become available \cite{santos}. We therefore compare our results
with those from \cite{santos} in the next few figures. Figure
\ref{kappa_p5} displays the change of aspect ratio of the ground
state for two extreme values of $\lambda=0.1, 10$ and for a small
value of $P=5$, and can be directly compared with the Fig. 3 of
\cite{santos}. We note that essentially the same results were
obtained as in \cite{santos} presumably because their neglect of
the s-wave interaction simply corresponds to $P=0$ of our more
general results.

\begin{figure}[h]
\centerline{\epsfig{file=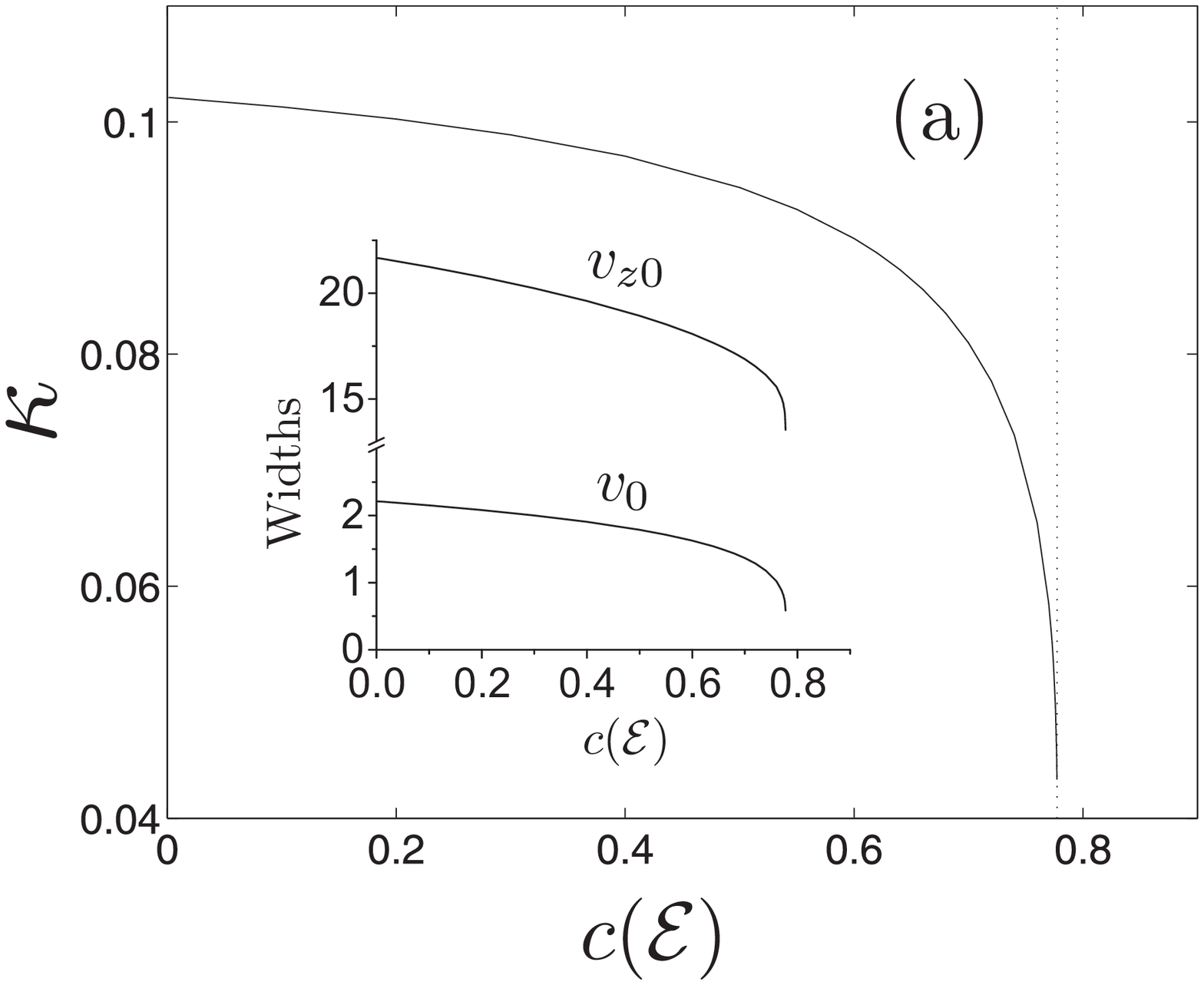,width=8.5cm}}
\centerline{\epsfig{file=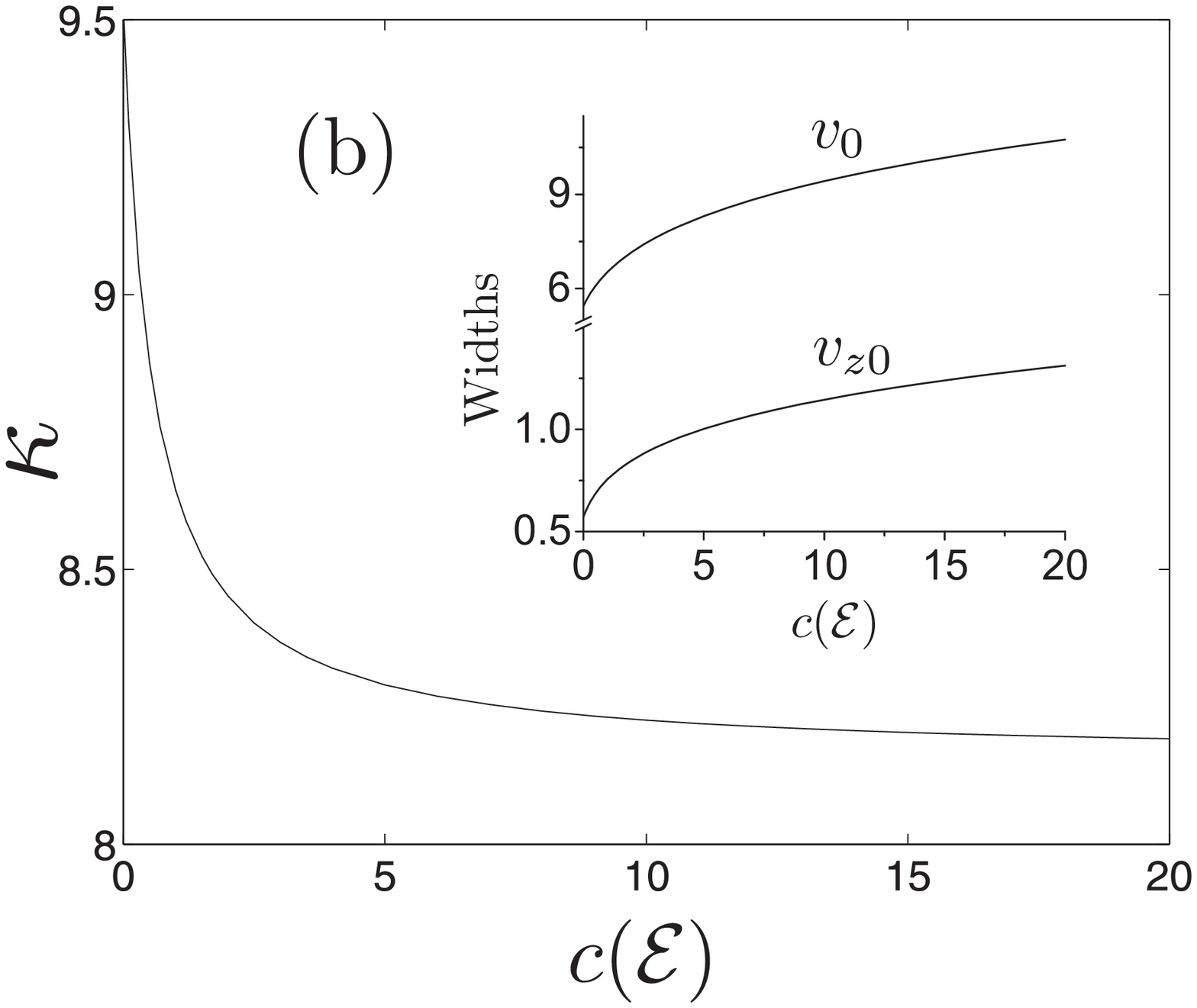,width=8.5cm}}
\caption{The same as in Fig. 14, but now for $P=500$.}
\label{kappa_p500}
\end{figure}

More interestingly, we show in Fig. \ref{kappa_p500} for a larger
value of $P=500$. We find the aspect ratio now changes in the
opposite direction with increasing dipole interaction strength.
This reversal of aspect ratio with increasing values of $P$ (due
to increasing in atom numbers or s-wave scattering length $a_{\rm
sc}$) is due precisely to the physics of minimizing the total
system free energy discussed earlier in the TFA. This phenomena
was not observed in the simpler model of Ref. \cite{santos}.  We
also find that $\lambda_c$ remains virtually independent of $P$
at the same value as in the TFA: 5.170169, consistent with Ref.
\cite{santos}.

\subsection{Evolution of widths} 
\begin{figure}[h]
\centerline{\epsfig{file=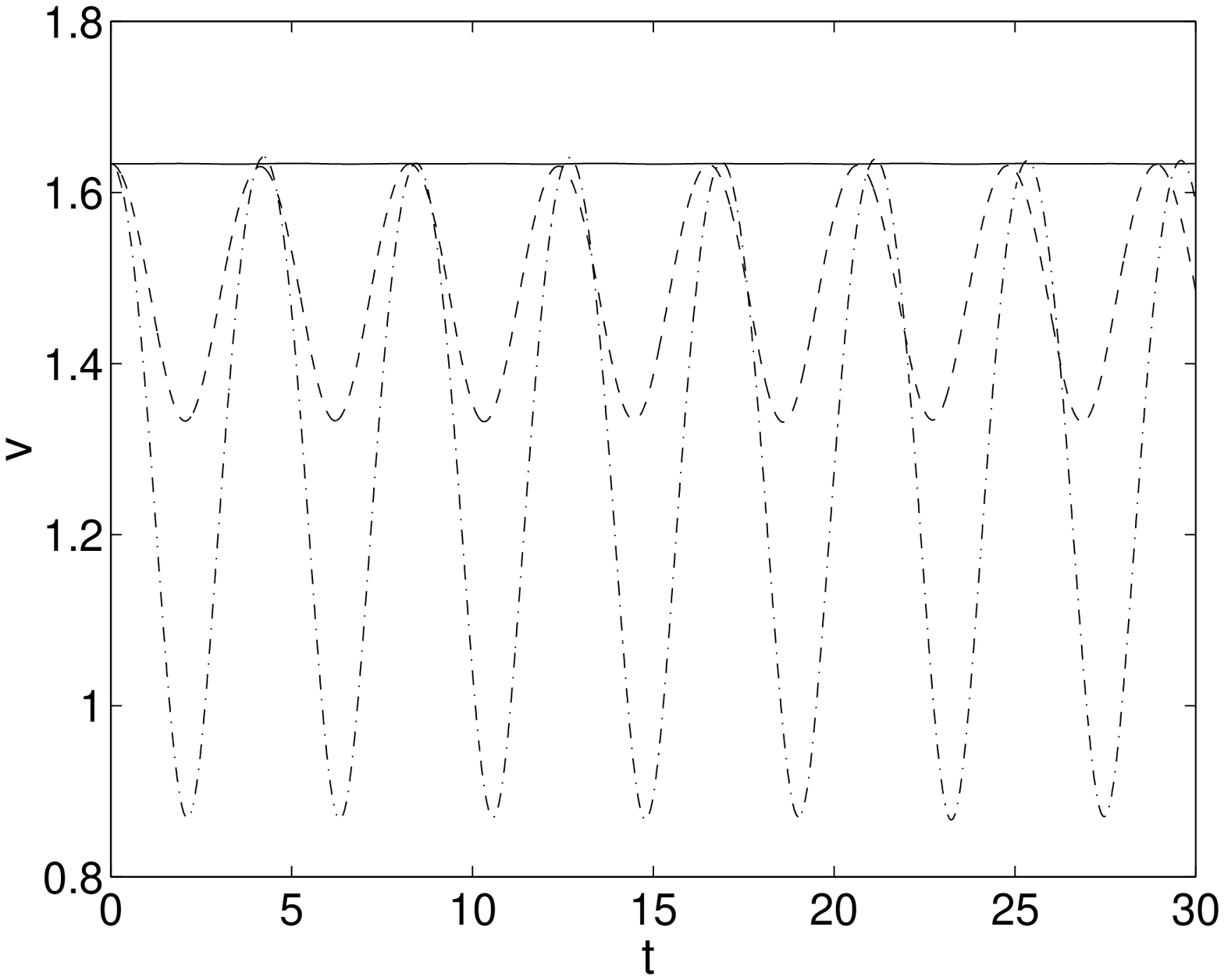,width=8.5cm}}
\centerline{\epsfig{file=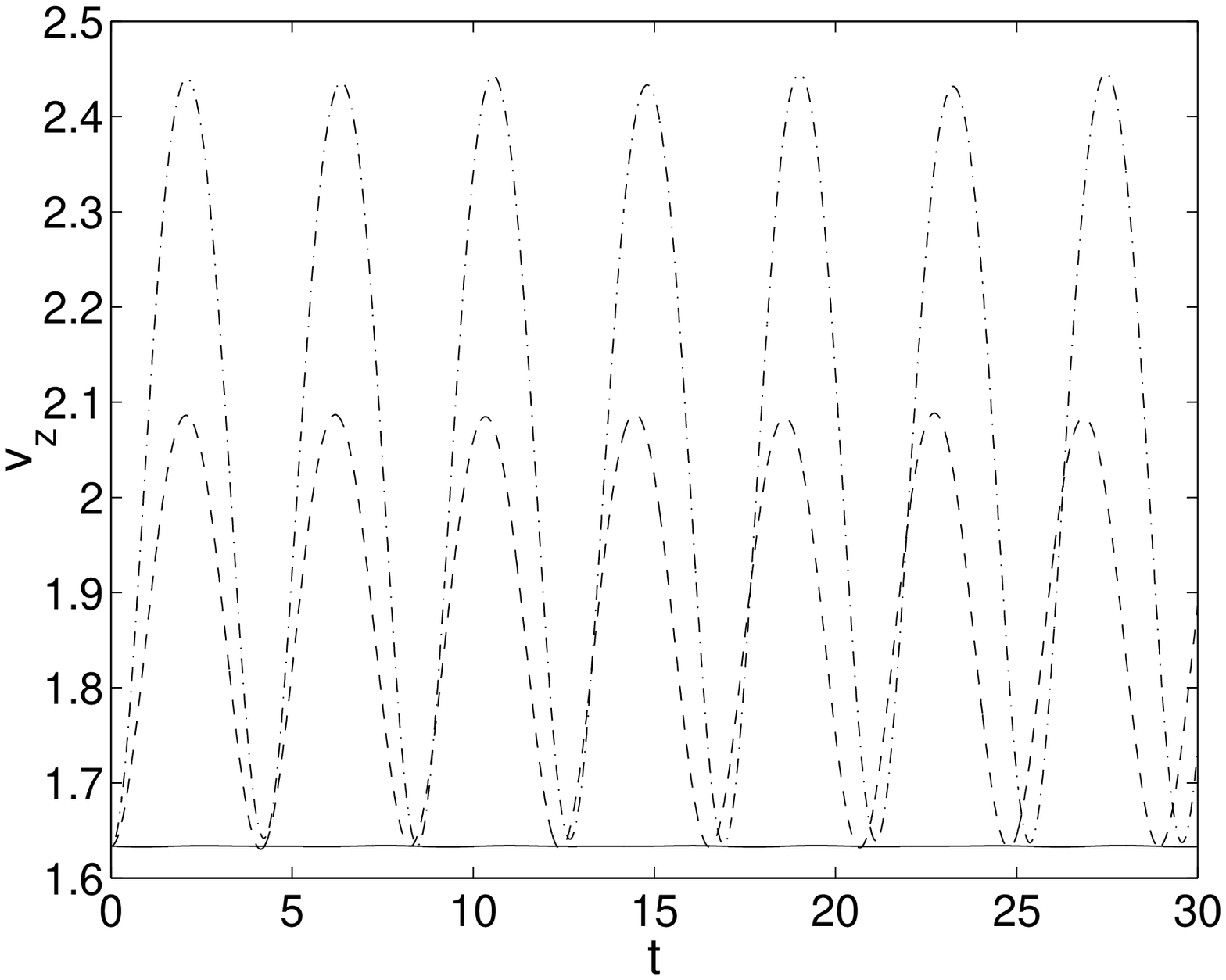,width=8.5cm}}
\caption{Evolution of condensate widths for $P=10$, $\lambda=1$,
$v(0)=v_z(0)=1.63359$ at $c({\cal E})=0.0$ (solid line), $c({\cal
E})=0.4$ (dashed line), and $c({\cal E})=0.8$ (dash-dotted
line).} \label{evowv}
\end{figure}

The evolution of condensate
widths are found by numerically integrating Eq. (\ref{evov1}).
Assuming initially $c({\cal E})=0$, $v(0)=v_0$, $v_z(0)=v_{z0}$,
and $\dot v(0)=\dot v_z(0)=0$, we can apply an electric field
suddenly or slowly for $t>0$. Under stable conditions for the
widths, we choose to apply electric field suddenly. Otherwise the
electric is increased linearly within a ramp-up time, and then
kept constant. First, for the stable case, Figure \ref{evowv}
shows condensate widths evolution up to $t=30$ (it has been
calculated up to $t=1000$). When $c({\cal E})=0$, the condensate
remains at its initially equilibrium state, and the widths are
unchanged with time. When $c({\cal E})\neq 0$, condensate widths
oscillate with time, and prolonged numerical propagation
indicates that we always have $v
>0$ and $v_z>0$, i.e. the condensate is stable.

\begin{figure}[h]
\centerline{\epsfig{file=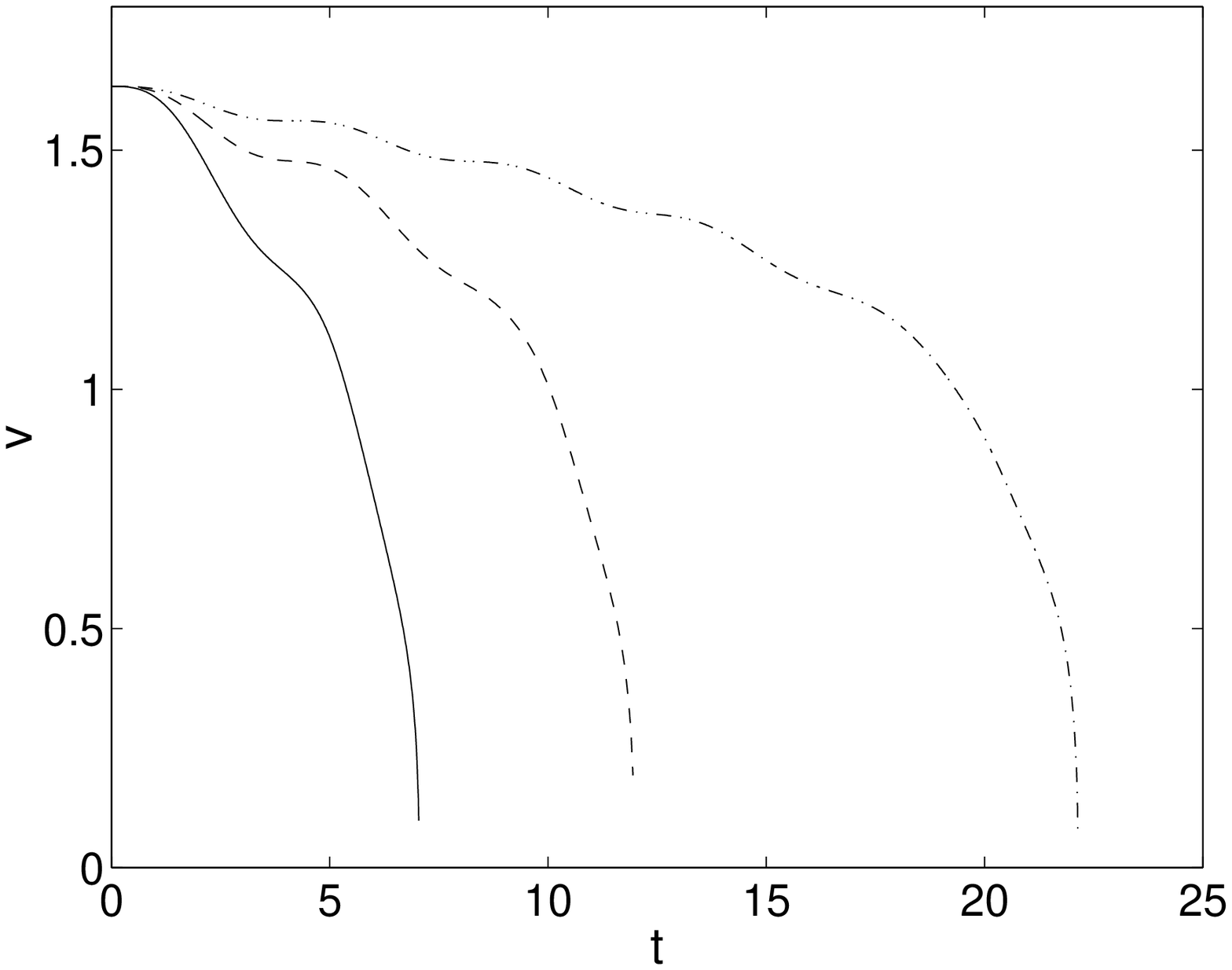,width=8.5cm}}
\centerline{\epsfig{file=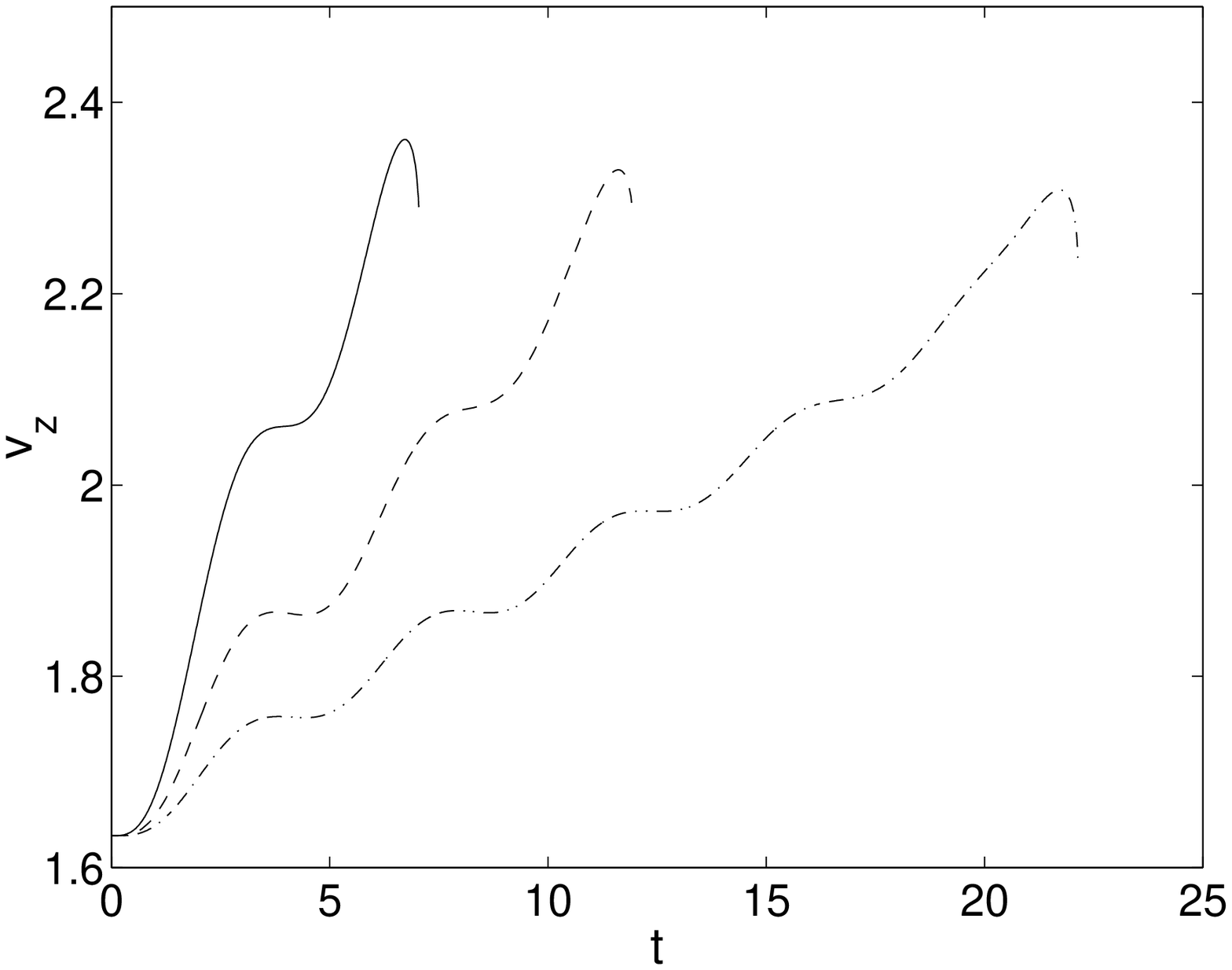,width=8.5cm}}
\caption{Evolution of condensate widths at $c({\cal E})=1.0$
($>c_M=0.9989$), for electric field ramp-up time $T=5$ (solid
line), $10$ (dashed line), and $20$ (dash-dotted lines). Other
parameters are $P=10$, $\lambda=1$, and $v(0)=v_z(0)=1.63359$. }
\label{evowcol}
\end{figure}

We also see from these figures that the oscillation amplitudes
increased with increasing $c({\cal E})$. Then finally at some
stage, we could arrive at $v<0$ or $v_z<0$, signaling the
condensate collapse. Figure \ref{evowcol} indeed displays such
cases when a linear ramp-on of the external electric field is
applied.

\subsection{Small amplitude shape oscillations} 
\begin{figure}[h]
\centerline{\epsfig{file=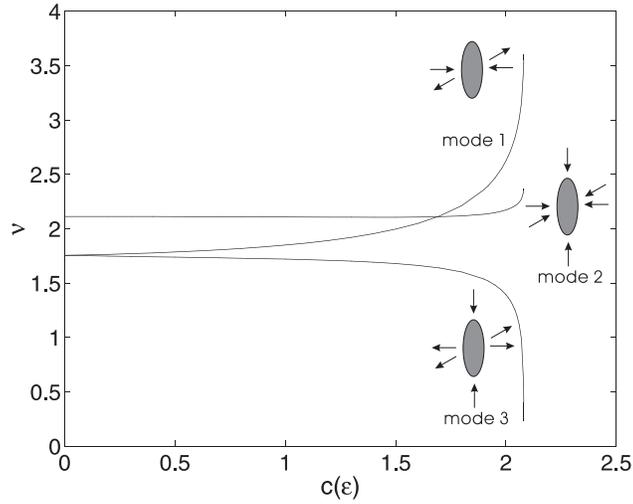,width=8.5cm}}
\caption{Electric field dependence of the shape oscillation
frequencies. Other parameters are $P=1$ and $\lambda=1$.}
\label{fig-modes}
\end{figure}

Once the
equilibrium widths are found from numerically solving Eqs.
(\ref{eqrootkv}), small amplitude oscillations can be studied by
evaluating the matrix of the second order derivatives of the
equivalent potential $U(v_x,v_y,v_z)$ Eq. (\ref{eqv}). We find
that it takes the following symmetric form
\begin{eqnarray}
\left(
\begin{array}{ccc} U_{11} & U_{12} & U_{13}\\ U_{12} & U_{11} & U_{13}
\\ U_{13} & U_{13} & U_{33}\end{array} \right),
\end{eqnarray}
where $U_{ij}=U_{ji}$ due to nature of commuting derivative
operations with different coordinates,
and $U_{11}=U_{22}$ and $U_{13}=U_{23}$ due to the
cylindrical symmetry. We find the oscillation
frequencies to be
\begin{eqnarray}
\nu_1 &&=\sqrt{U_{11}-U_{12}}, \nonumber\\ \nu_{2,3} &&={1\over
\sqrt 2}\left[U_{11}+U_{12}+U_{33}\pm\sqrt{U^2_{11}+U^2_{12}
+U^2_{33}+8U^2_{13}+2U_{11}U_{12}-2U_{11}U_{33}-2U_{12}U_{33}}\right]^{1/2},
\end{eqnarray}
where the expression for the matrix elements $U_{ij}$ are listed
in Appendix \ref{appb}. Typical results and mode structure
identifications \cite{var1} are given in Fig. \ref{fig-modes}. We
see that mode 1 and mode 3 are doubly degenerate when $c({\cal
E})=0$. This is due to the additional symmetry $U_{11}=U_{33}$
and $U_{12}=U_{13}$ for $\lambda=1$.

\section{conclusion}
In conclusion, we have performed a detailed
study of trapped condensates with dipole interactions.
We have developed a general scheme for
constructing effective pseudo-potentials for anisotropic
interactions \cite{ring}, which guarantees the agreement between
the first order Born scattering amplitude from
the pseudo-potential and the complete
scattering amplitude obtained from a multi-channel
collision calculation.
Our theory has been applied to the study of
induced electric dipole interactions and can also be
directly extended to magnetic dipole interactions
as well as permanent electric dipole interactions
of trapped molecules \cite{doyle}.

Finally we provide a reality check for prospects of
experimental observations of the electric field induced
interaction effects. Though the required fields
are relatively high, there are evidences they can be
created with current laboratory technology.
In Ref. \cite{Meijer} fields of upto $2\times 10^5$ (V/cm)
were used to slow a molecular beam.
Gould \cite{Gould} used fields
upto $4.6\times 10^5$ (V/cm) in the measurement of
atomic tensor polarizability, while Marrus {\it et al.} \cite{Marrus}
reported fields upto $10^6$ (V/cm).
What is perhaps most encouraging is a recent
experiment for cooling molecule beams with time-dependent
(adiabatic from the view point of atomic internal dynamics)
fields of upto $10^7$ (V/cm) \cite{Maddi}.
We also note that at the high fields being discussed in
this paper, the tunneling ionization of atoms remain
infinitesimally small \cite{Robin}.

This work is supported by the U.S. Office of Naval Research grant
No. 14-97-1-0633 and by the NSF grant No. PHY-9722410. The
computation of this work is partially supported by NSF through a
grant for the ITAMP at Harvard University and Smithsonian
Astrophysical Observatory.

\appendix

\section{U-matrix elements}
\label{appb}
After tedious calculations, we find that
\begin{eqnarray}
U_{11}=1+{3\over v^4_0}+{P\over
v^4_0v_{z0}}\left[2-{\sqrt{5\pi}c({\cal E})\over
24(v^2_{z0}-v^2_0)^3}\left(32v^6_0+141v^4_0v^2_{z0}
-54v^2_0v^4_{z0}+16v^6_{z0}-9(11v^2_0+4v^2_{z0})v^4_0H(v_0/v_{z0})\right)\right],
\end{eqnarray}
\begin{eqnarray}
U_{33}=\lambda^2+{3\over v^4_{z0}}+{P\over
v^2_0v^3_{z0}}\left[2-{\sqrt{5\pi}c({\cal E})\over
3(v^2_{z0}-v^2_0)^3}\left(4v^6_0-12v^4_0v^2_{z0}
+51v^2_0v^4_{z0}+2v^6_{z0}-9(v^2_0+4v^2_{z0})v^2_0v^2_{z0}H(v_0/v_{z0})\right)\right],
\end{eqnarray}
\begin{eqnarray}
U_{12}={P\over v^4_0v_{z0}}\left[1-{\sqrt{5\pi}c({\cal E})\over
24(v^2_{z0}-v^2_0)^3}\left(16v^6_0+51v^4_0v^2_{z0}
-30v^2_0v^4_{z0}+8v^6_{z0}-45v^6_0H(v_0/v_{z0})\right)\right],
\end{eqnarray}
and
\begin{eqnarray}
U_{13}={P\over v^3_0v^2_{z0}}\left[1-{\sqrt{5\pi}c({\cal E})\over
6(v^2_{z0}-v^2_0)^3}\left(4v^6_0-36v^4_0v^2_{z0}
-15v^2_0v^4_{z0}+2v^6_{z0}+45v^4_0v^2_{z0}H(v_0/v_{z0})\right)\right].
\end{eqnarray}

\end{document}